\documentclass[twocolumn]{aastex631}
\usepackage{listings}
\usepackage{hyperref}
\usepackage{threeparttable}

\shorttitle{CHIPS Open Source Code}
\shortauthors{Takei et al.}

\begin{document}

\title{CHIPS: Complete History of Interaction-Powered Supernovae}

\author{Yuki Takei}
\affiliation{Research Center for the Early Universe (RESCEU), School of Science, The University of Tokyo, 7-3-1 Hongo, Bunkyo-ku, Tokyo 113-0033, Japan}
\affiliation{Department of Astronomy, School of Science, The University of Tokyo, Tokyo, Japan}
\affiliation{Astrophysical Big Bang Laboratory, RIKEN, 2-1 Hirosawa, Wako, Saitama 351-0198, Japan}

\author{Daichi Tsuna}
\affiliation{Research Center for the Early Universe (RESCEU), School of Science, The University of Tokyo, 7-3-1 Hongo, Bunkyo-ku, Tokyo 113-0033, Japan}
\affiliation{Department of Physics, School of Science, The University of Tokyo, Tokyo, Japan}

\author{Naoto Kuriyama}
\affiliation{Research Center for the Early Universe (RESCEU), School of Science, The University of Tokyo, 7-3-1 Hongo, Bunkyo-ku, Tokyo 113-0033, Japan}
\affiliation{Department of Astronomy, School of Science, The University of Tokyo, Tokyo, Japan}

\author{Takatoshi Ko}
\affiliation{Research Center for the Early Universe (RESCEU), School of Science, The University of Tokyo, 7-3-1 Hongo, Bunkyo-ku, Tokyo 113-0033, Japan}
\affiliation{Department of Astronomy, School of Science, The University of Tokyo, Tokyo, Japan}

\author{Toshikazu Shigeyama}
\affiliation{Research Center for the Early Universe (RESCEU), School of Science, The University of Tokyo, 7-3-1 Hongo, Bunkyo-ku, Tokyo 113-0033, Japan}
\affiliation{Department of Astronomy, School of Science, The University of Tokyo, Tokyo, Japan}

\correspondingauthor{Yuki Takei}
\email{takei@resceu.s.u-tokyo.ac.jp}



\begin{abstract}
We present the public release of the Complete History of Interaction-Powered Supernovae (CHIPS) code, suited to model a variety of transients that arise from interaction with a dense circumstellar medium (CSM). Contrary to existing modellings which mostly attach the CSM by hand, CHIPS self-consistently simulates both the creation of the CSM from mass eruption of massive stars prior to core-collapse, and the subsequent supernova light curve. We demonstrate the performance of CHIPS by presenting examples of the density profiles of the CSM and the light curves. We show that the gross light curve properties of putative interaction-powered transients, such as Type IIn supernovae, rapidly evolving transients and recently discovered fast blue optical transients, can be comprehensively explained with the output of CHIPS.
\end{abstract}


\section{Introduction}
 {Core-collapse} supernovae (SNe) are explosive events which occur at the end of massive stars' lives.
They show a variety of features in their spectra, mainly depending on the chemical abundances of the progenitor.
Though emission lines are usually broad, some SNe show very narrow hydrogen emission lines in their spectra, which are classified as Type IIn \citep{1990MNRAS.244..269S}.
The narrow width strongly suggests that the progenitor is surrounded by a slow-moving dense circumstellar medium (CSM). A dense CSM also efficiently converts the kinetic energy of the SN ejecta to radiation. Such ``interaction-powered" events often have brighter light curves (LCs) than other types of SNe powered by the internal energy and/or radioactive decay of $^{56}\mathrm{Ni}$ and $^{56}\mathrm{Co}$ in the ejecta \citep{Richardson14}.

Recent high cadence transient surveys have found that SNe IIn are not the entirety of interaction-powered transients. These surveys have discovered luminous transients that display rapid rise and decline (of within 10 days) in the LC (e.g. \citealt{Drout_et_al_2014,Arcavi_et_al_2016,Pursiainen_et_al_2018}), which are sometimes classed as rapidly evolving transients (RETs). Furthermore, in the past few years there have been a handful of discoveries of transients with even faster evolution on the order of days and luminosity exceeding $10^{44}\ {\rm erg\ s^{-1}}$, named as fast blue optical transients (FBOTs; e.g., \citealt{Prentice18,Perley19,Ho20,Perley21}). While multiple interpretations exist for these transients, a fraction of these transients also show signs of the presence of dense CSM, that can potentially explain the brightness of these transients as well.

The key questions for these interaction-powered SNe are how the dense CSM has formed, and how the CSM is distributed around the SN progenitor when it explodes.
In general, line emissions of metal induce steady mass loss with a rate up to $\sim10^{-3}\,M_\odot\,{\rm yr^{-1}}$ \citep[e.g.,][]{1975ApJ...195..157C,2001A&A...369..574V}.
On the other hand, \citet{Kiewe2012} estimate from the strength of the H$\alpha$ emission that the mass-loss rates of the progenitor of SNe IIn reach $\sim0.1M_\odot\,{\rm yr^{-1}}$, which cannot be achieved by the line-driven mass loss.
While theoretical models suggest some mechanisms to transport energy which induce eruptive mass loss from massive stars \citep[e.g.,][]{2007Natur.450..390W,2012MNRAS.423L..92Q, 2014A&A...564A..83M, 2014ApJ...785...82S, 2015ApJ...810...34W, 2017MNRAS.464.3249S}, there is no observational constraint on how they lose their envelope intensely. By contrast, there are recent analyses of SN observations which show that some massive stars experience eruptive mass loss before the actual SN event to form CSM, the famous example being SN 2009ip \citep[e.g.,][]{2004MNRAS.352.1213C,2007Natur.447..829P,2013ApJ...767....1P,Ofek14,Elias-Rosa18}.

Since interaction-powered SNe shine by converting the kinetic energy into thermal energy through interaction between SN ejecta and dense CSM, their LCs strongly depend on the density structure of CSM.
%The LCs of interacting supernovae are thus direct probes for the surrounding CSM.
%Therefore, theoretical investigation into the LC leads to the prediction of the activity of the progenitor just before the core-collapse, too faint compared to a SN event to be observed.
Thus, the theoretical investigation into the LCs of interacting supernovae are direct probes for the surrounding CSM as well as the progenitor activity just before the core-collapse, which is too faint compared to an SN event to be observed.

Many works modelling LCs of interaction-powered SNe concentrate on the CSM formed by steady mass-loss $\rho\propto r^{-2}$ (where $\rho,\,r$ denote the density and radius), or CSM which follows a single power-law \citep[e.g.,][]{2011MNRAS.415..199M,2012ApJ...757..178G,2012ApJ...746..121C,Moriya_et_al_13, Dessart15, Tsuna19, Takei20, Suzuki20}. However, it is unclear whether an eruptive mass loss event makes such a simple density structure around a star.

Recent numerical simulations indicate that the energy deposition at the stellar envelope likely triggers the mass eruption, although the precise energy transport mechanism is unknown \citep{Dessart_et_al_2010, Owocki_et_al_19, Kuriyama20a}. In particular, \citet{Kuriyama20a} conducted one-dimensional radiation hydrodynamics simulations by injecting energy comparable to the binding energy of the envelope. They not only successfully reproduced the mass of the CSM inferred from Type IIn SNe, but also found that the density profile is quite different from a simple power-law. Hence, numerical simulations for eruptive mass loss is necessary in order to accurately predict the observations of interaction-powered SNe from LC modelling.

For better modelling of interaction-powered SNe, we have developed the open source code ``Complete History of Interaction-Powered Supernovae (CHIPS)"\footnote{\href{https://github.com/DTsuna/CHIPS}{https://github.com/DTsuna/CHIPS}. Version 1.0.1 is archived in Zenodo \citep{CHIPS}} that self-consistently calculates the LC of SNe interacting with the mass eruption from the progenitor.
We show that the output of CHIPS comprehensively reproduces the LC properties of the observed interaction-powered transients. Here we introduce the code in detail and show some representative results. Should an interesting interaction-powered transient be discovered in the future, CHIPS would serve as a code for theoretical modelling of the LC and CSM.

This paper is constructed as follows. In Section \ref{sec:overview} we briefly outline the calculations that are done in the CHIPS code, and what can be obtained as outputs of this code. In Section \ref{sec:technical_details} we describe the technical details of the CHIPS code, focusing on the formulations and underlying assumptions of the mass eruption and LC calculations. In Section \ref{sec:Results} we show results that we obtained from a series of calculations performed with the CHIPS code, and explore the variety of CSM density profiles and LCs obtained from our parameter study. We discuss the planned future updates of the code in Section \ref{sec:future_work}, and conclude in Section \ref{sec:conclusion}. Appendix A summarizes how to install and execute CHIPS code and what kind of results are obtained.

\section{Overview}
\label{sec:overview}

 \begin{figure*}
 \centering
 \includegraphics[width=\linewidth]{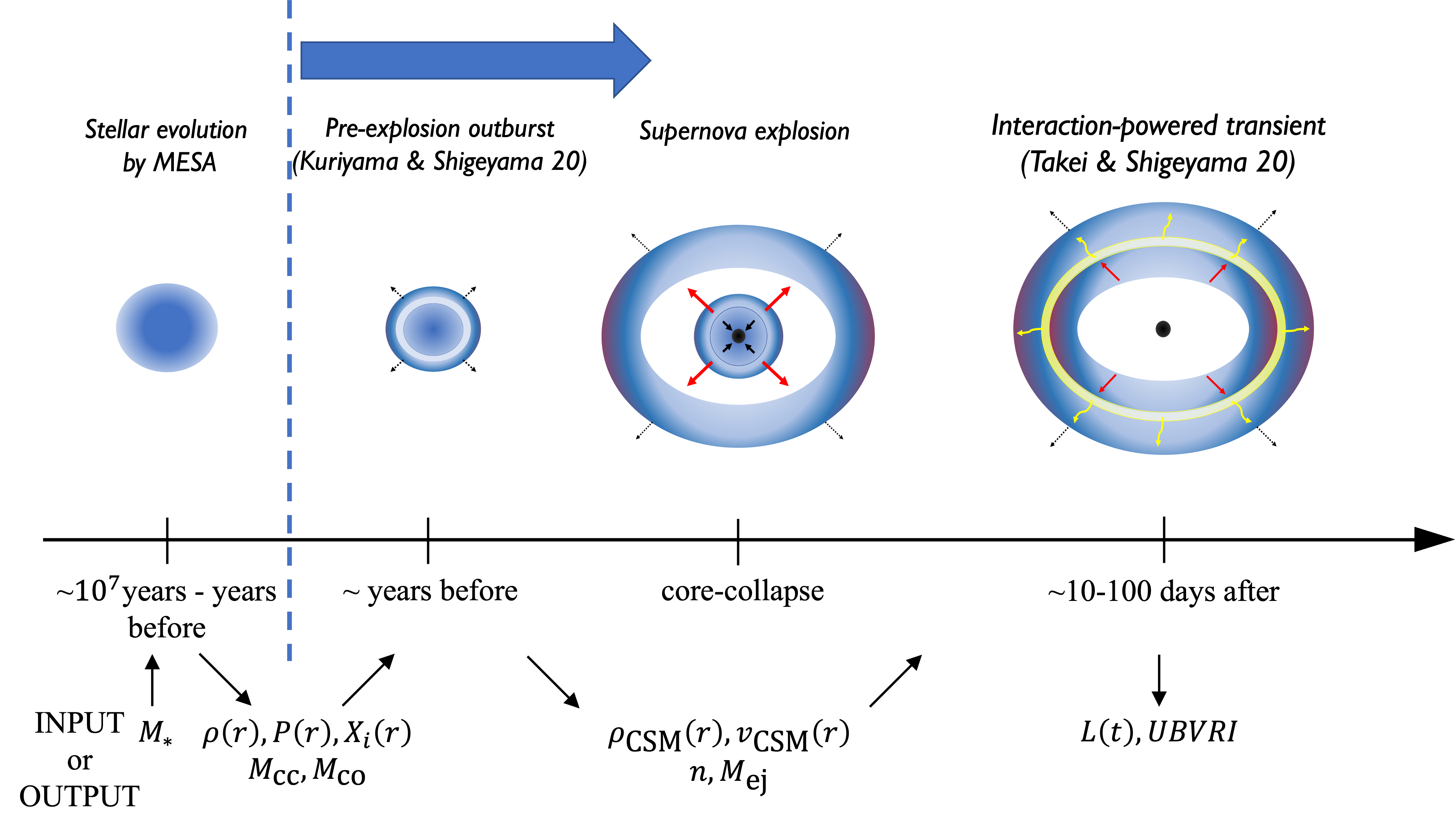}
\caption{Schematic overview  {and flowchart }of the CHIPS code. The code first simulates the outburst just prior to core-collapse that creates a dense CSM. It then simulates the light curve from a supernova explosion powered by interaction with this CSM.}
 \label{fig:schematic}
 \end{figure*}

We show the schematic picture of our code in Figure \ref{fig:schematic}. The CHIPS code is composed of the following three components:
\begin{enumerate}
\item The code first loads a stellar model obtained by the MESA stellar evolution code \citep{Paxton11,Paxton13,Paxton15,Paxton18,Paxton19}. The user can generate their own model using MESA, or use a sample stellar model at core-collapse generated by the authors using the \verb|example_make_pre_ccsn| test suite in MESA version 12778. We have verified that the code is compatible with output files generated by the latest revision of MESA (r15140) at the time of writing.
\item Then using the input stellar model, the mass eruption and the resulting CSM are calculated using a radiation hydrodynamical code based on the code developed in \cite{Kuriyama20a}, as detailed in Section \ref{sec:mass_eruption}.
\item Finally the bolometric and multi-wavelength LCs powered by the CSM interaction are numerically calculated using the methods detailed in Section \ref{sec:light_curve}. This calculation is based on the code developed in \cite{Takei20}.
\end{enumerate}
For the first step, the stellar model can be specified with an appropriate path using the argument \verb|--stellar-model| in the execution script \verb|run.py|. Our sample consists of stars in a ZAMS mass range of $13-26M_\odot$ and with ZAMS metallicity of solar, and is used for the demonstration of our code in Section \ref{sec:Results}.

For historical reasons the code for calculating the mass eruption is written in Fortran, and that for obtaining the LC is written in C. These codes can be compiled and executed with the standard gfortran and gcc compilers. The execution script \verb|run.py| also internally does computationally inexpensive calculations to connect these steps. These include remeshing radial profiles of the star and CSM, generating opacity tables, and obtaining the gross property of the emission (e.g. peak luminosity, rise time) from the LC data.

Apart from various physical parameters that can be set in the stellar evolution done in MESA, our code consists of four parameters as shown in Table \ref{table:Parameters}. We also show the range of values recommended by the authors to get a physical and reliable result. The detailed procedures to install and run the CHIPS code is shown in Appendix \ref{sec:CHIPS_execution}.

\begin{table*}
\centering
\begin{threeparttable}
\caption{The key parameters in the CHIPS code that the user can tune. In the last column we show the range of values recommended by the authors.}
\begin{tabular}{c|cc}
\hline\hline
Parameter & Definition & Recommended values \\ \hline
$M_*$ & Mass of star at ZAMS & $13$--$26$ $M_\odot$ \\
$f_{\rm inj}$ & Injected energy nomalized by the envelope's binding energy & $0.3$--$0.8$\tnote{$a$}\\
$t_{\rm inj}$ & Time from energy injection to core-collapse & $3$--$30$ years\tnote{$a$}\\
$E_{\rm ej}$ & Explosion energy of supernova ejecta & $3\times 10^{50}$--$10^{52}$ ergs\tnote{$b$} \\ \hline
\end{tabular}
\begin{tablenotes}\footnotesize
\item[$a$]  {This parameter space reproduces the general characteristics (mass, radii) of the CSM observed in Type IIn supernovae \citep{Kuriyama20a}.}
\item[$b$]  {%Although this range covers explosions of normal SNe and hypernovae, we cannot simulate weak explosions (see also Section \ref{sec:parameter}).
Our assumption of homologous ejecta limits the explosion energy to sufficiently high energies (see also Section \ref{sec:parameter}).}
\end{tablenotes}
\label{table:Parameters}
\end{threeparttable}
\end{table*}

\section{Description of the CHIPS Code}
\label{sec:technical_details}
\subsection{Mass Eruption Calculation}
\label{sec:mass_eruption}
The dynamics of the mass eruption and the formation of CSM are calculated by our 1-D Lagrangian radiation  hydrodynamical simulation code developed by \citet{Kuriyama20a} and \citet{Kuriyama21}. Our code adopts the stellar model generated by the MESA code as the initial condition. The hydrogen-rich envelope is extracted from this file as the calculation target, and remeshed in order to enhance the resolution and to reduce the differences of the cell masses between neighboring cells prior to our hydrodynamical calculation. The total number of cells is specified by a parameter \verb|hydroNumMesh|, with a default value of 10000.

%\begin{figure*}
% \centering
% \includegraphics[width=\linewidth]{preccsn_15Msun.pdf}
%\caption{ {Left panel: Mass fraction profiles of $M_*=15M_\odot$ just before core-collapse. The mass of helium core is $4.9M_\odot$. Right panel: Density and temperature profiles as functions of the enclosed mass.}}
% \label{fig:preccsn}
%\end{figure*}

After a short (50 seconds) initial relaxation of the remeshed initial models, thermal energy is injected into the inner boundary region, and the subsequent dynamical eruption is simulated. If the progenitor has a sufficiently massive ($>0.2M_\odot$) hydrogen envelope, the default mass coordinate of the inner boundary is set to be $0.2M_\odot$ outside its helium core  {to avoid a severe Courant condition caused by a sudden increase of the density around the boundary between the helium core and hydrogen envelope.} Otherwise, a custom inner boundary must be set by the argument \verb|--eruption-innerMr| (in units of solar masses) when executing \verb|run.py|. This code is agnostic of the energy source, but the amount of injected energy and the duration of the injection are set by the parameters $f_{\rm inj}$ and \verb|injectDuration| respectively. The parameter $f_{\rm inj}$ is defined as the ratio of the injected energy and the absolute value of the total energy inside the envelope
\begin{equation}
    E_{\rm tot}=\int 4\pi r^2\rho dr \left(\frac{\upsilon^2}{2}+e_{\rm int}-\frac{GM_r}{r}\right),
\end{equation}
where $r,v,\rho,e_{\rm int},M_r$ are respectively the distance from the center, velocity, density, specific internal energy and enclosed mass, $G$ is the gravitational constant, and the three terms in the parenthesis are respectively the kinetic energy, internal energy and gravitational energy. Since the progenitor's envelope is gravitationally bound, $E_{\rm tot}$ is negative for all progenitors. \citet{Kuriyama20a} finds that $f_{\rm inj}$ of a few 10\% results in partial ejection of the envelope and that the total ejected mass is sensitive to the value of $f_{\rm inj}$.

The parameter \verb|injectDuration| is by default set to 1000 seconds, which is much shorter than the dynamical timescale in the envelope
\begin{eqnarray}
 t_{\rm dyn} \sim \sqrt{\frac{R_*^3}{GM_*}} \sim 10^7\ {\rm sec} \left(\frac{R_*}{5\times 10^{13}\ {\rm cm}}\right)^{3/2}\left(\frac{M_*}{10M_\odot}\right)^{-1/2},
\end{eqnarray}
where $R_*$ and $M_*$ are respectively the radius and mass of the progenitor. A longer timescale comparable to $t_{\rm dyn}$ would make the gravitational pull from the central core important, thereby reducing the mass and energy of the erupted material  {\citep{Ko21}.} %We are exploring this dependence in an independent study (Ko et al., in preparation).

A part of the envelope is erupted on the dynamical timescale and becomes a dense CSM. After enough time ($\gtrsim 2t_\mathrm{dyn}$) has passed since the mass eruption, the ejected envelope expands and radiation pressure gradient can no longer affect its dynamics. Therefore, at this stage the radiative transfer calculation scheme is turned off to save computational cost. In addition, because our interest is on the CSM profile at core collapse, the innermost region ($<10^{13}$ cm) of the envelope is excluded from the computational range. The hydrodynamical simulation is continued until core-collapse, with duration given by the parameter $t_{\rm inj}$ (see Table \ref{table:Parameters}).

\begin{figure}
 \centering
 \includegraphics[width=\linewidth]{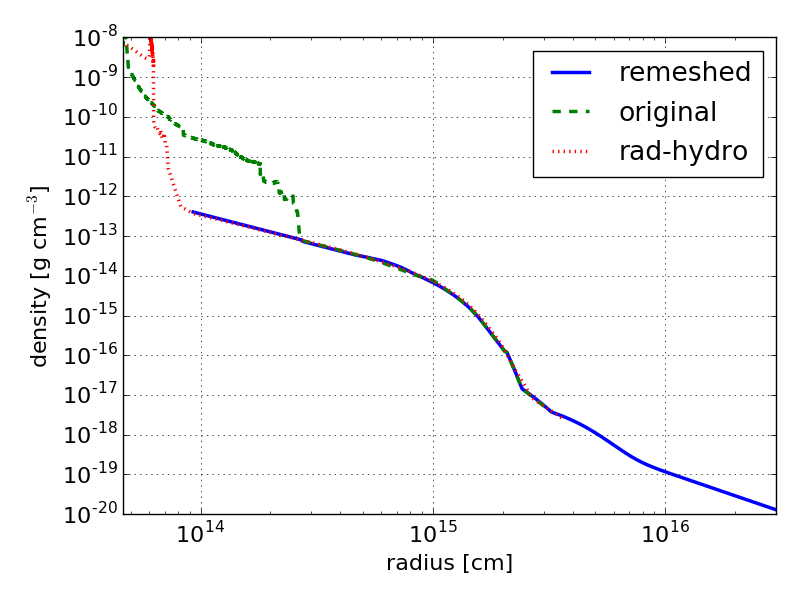}
\caption{Example of remeshing of the CSM density profile done in our code. The solid line is reconstructed from the output of a hydrodynamical simulation (dashed line) using equation (\ref{eq:rho_CSM}). The solid line matches the profile calculated from a more rigorous radiation hydrodynamical simulation (dotted line). The profiles are for energy injection into a star with ZAMS mass of $15M_\odot$, with parameters $f_{\rm inj}=0.3$ and $t_{\rm inj}=10$ years. This figure (except for the dotted line) is automatically generated by the code when remeshing of the CSM is done.}
 \label{fig:csmremesh}
 \end{figure}

%ここ直す?
The density profile created after turning off radiative transfer is found to contain artificial shocks that propagate to the CSM (see green dashed line of Figure \ref{fig:csmremesh}). This is because part of the bound CSM falls back from the star, creating shocks that do not disappear in the adiabatic approximation. Since the cooling and diffusion timescales in the shock-heated region are much shorter than the age, we expect that the radiative cooling will suppress these discontinuities from propagating to the CSM if correctly taking into account radiative transfer. In other words the discontinuity is artificial, and the profile should instead smoothly connect to the region unaffected by the shock. This is verified from long-term radiation hydrodynamical simulations done in \citet{Tsuna21b} as seen in the red dotted line of the same figure. 

To correctly take into account this effect, the code updates the density profile with an analytical model in \citet{Tsuna21b},
\begin{eqnarray}
\rho_{\rm CSM}(r) = \rho_* \left[\frac{(r/r_*)^{1.5/y}+(r/r_*)^{12/y}}{2}\right]^{(-y)}.
\label{eq:rho_CSM}
\end{eqnarray}
where $r_*$, $\rho_*$, $y$ are the fitting parameters.
The code fits the outer region unaffected by the shock to the analytical profile, and extrapolates the fit inwards to $r_\mathrm{in}=2R_*$, where $r_\mathrm{in}$ denotes the innermost radius of the CSM. This choice of the inner boundary is rather arbitrary; in our case it is mainly from the assumption of homologous supernova ejecta, which is valid only after the ejecta expands roughly twice its initial radius.  {Although $r_\mathrm{in}$ is dependent upon $M_*,\,f_\mathrm{inj}$, it does not affect the calculations of the CSM mass and thus the subsequent LCs (see next section).}

Outside the erupted material, the code adds by hand CSM composed of a stellar wind emitted by the progenitor before the eruptive mass loss. The CSM is assumed to have a steady wind profile $\rho=\dot{M}/4\pi r^2 \upsilon_w$ that extends out to $3\times 10^{16}$ cm. The choice of the outer radius does not affect the final LC since the optical depth is only determined by the inner CSM. The mass loss rate $\dot{M}$ and wind velocity $\upsilon_w$ are assumed to scale with the progenitor's luminosity $L_*$ and effective temperature $T_*$ as \citep{Nieuwenhuijzen90,Mauron11}\footnote{We checked whether the output of the mass-loss rate from MESA is consistent with the value obtained from this equation, and confirmed that the mass-loss rates well match with a difference of at most a factor of 2.}
\begin{eqnarray}
 \dot{M} &=& 5.6\times 10^{-6}M_\odot{\rm yr}^{-1} \left(\frac{L_*}{10^5L_\odot}\right)^{1.64} \left(\frac{T_*}{3500\ {\rm K}}\right)^{-1.61},\label{eqn:mass_loss_from_mesa}\\
 \upsilon_w &=& 20\ {\rm km\ s^{-1}} \left(\frac{L_*}{10^5L_\odot}\right)^{0.35}.
\end{eqnarray}
The values $L_*$ and $T_*$ are taken from the stellar model generated by MESA. This prescription is purely for extending the computational region. Since the mass carried by the stellar wind is much smaller than the mass of the erupted material, the details of this do not affect the major characteristics of the final LC. A demonstration of this remeshing process is shown in Figure \ref{fig:csmremesh}. The correction of artificial shocks in the inner region successfully reproduces the profile calculated from long-term radiation hydrodynamical simulations.

\subsection{Light Curve Calculation}
\label{sec:light_curve}
Several years to decades after the mass eruption, a supernova explosion occurs and the ejecta collide with the CSM. The shocked region formed from circumstellar interaction is enclosed with the reverse and forward shocks, which propagates into the SN ejecta and the unshocked CSM, respectively.

The radiative flux emerging from the forward shock heats the unshocked CSM. The CHIPS code calculates the bolometric and multi-band LCs following the methodology in \citet{Takei20}. The details of the LC model are mostly the same as \citet{Takei20}, but we have made several important updates for performance and compatibility with CHIPS. Thus we describe the detailed setup and the numerical procedures, as well as various updates from \citet{Takei20}.

\subsubsection{Parameter setup}\label{sec:parameter}
For the SN ejecta, we adopt homologous ejecta with a double power-law density profile  \citep{Matzner99}
\begin{eqnarray}
\rho_{\rm ej} (r,t)
&=& \left\{ \begin{array}{ll}
t^{-3}\left[r/(gt)\right]^{-n} & (r/t > \upsilon_t),\\
t^{-3}(\upsilon_t/g)^{-n} \left[r/(t\upsilon_t)\right]^{-\delta}  & (r/t < \upsilon_t),
\end{array}\right.
\label{eq:rho_ej}
\end{eqnarray}
which is valid roughly after the ejecta expand to $r\approx 2R_*$ and kinetic energy dominates over thermal energy. The constants $g$ and $\upsilon_t$ are functions of the ejecta mass $M_{\rm ej}$ and energy $E_{\rm ej}$ as
\begin{eqnarray}\label{eq:coeff_ej}
g &=& \left\{\frac{1}{4\pi(n-\delta)} \frac{[2(5-\delta)(n-5)E_{\rm ej}]^{(n-3)/2}}{[(3-\delta)(n-3)M_{\rm ej}]^{(n-5)/2}}\right\}^{1/n},\\
\upsilon_t &=& \left[\frac{2(5-\delta)(n-5)E_{\rm ej}}{(3-\delta)(n-3)M_{\rm ej}}\right]^{1/2}.\label{eq:v_t}
\end{eqnarray}
The two exponents $n$ and $\delta$ are constant. The explosion energy $E_{\rm ej}$ is a free parameter in the code. The code calculates three light curves for the three different energies $\{1,3,10\}\times 10^{51}$ erg by default. The ejecta mass $M_{\rm ej}$ is calculated from
\begin{equation}
    M_{\rm ej} = M_{\rm cc} - M_{\rm CSM} - M_{\rm rem}(M_{\rm CO})
\end{equation}
where $M_{\rm cc}$ is the mass of the star at core-collapse obtained from the MESA stellar model, $M_{\rm CSM}$ is the mass of the CSM measured from $r=2R_*$, and $M_{\rm rem}$ is the mass of the remnant compact object formed in the center.
%半径分からないから2R*くらいまで膨らんだと仮定したと書いて、エジェクタに質量いってるとする
 {Since we cannot find the true radius of the progenitor due to the artificial shock as shown in Figure \ref{fig:csmremesh}, we assume that the progenitor expands to $r=2R_*$ at core-collapse\footnote{ {The true radius at core-collapse depends on other parameters of mass eruption such as $t_\mathrm{inj},\,f_\mathrm{inj},\,M_*$, having a range of $(1-2)R_*$ \citep[][Figures 1,\,3]{Tsuna21b}.}}. Thus the mass enclosed between the true radius and $r=2R_*$ is contained in the ejecta.}
 {From equation (\ref{eq:rho_CSM}), the mass of the CSM is estimated to be $M_\mathrm{CSM}\simeq\int_{2R_*}^{r_*}4\pi r^{2}\rho_\mathrm{CSM}(r)dr\propto r_{*}^{1.5}-(2R_*)^{1.5}$ and very weakly depends on the choice of the inner boundary as long as $2R_* \ll r_*$. This is generally satisfied for $t_{\rm inj}$ of a few years or longer.}
The value of $M_{\rm rem}$ is estimated from the mass of the progenitor's carbon-oxygen core $M_{\rm CO}$ (also obtained from the MESA model), using a simple formula obtained for single stars \citep{Schneider21}
\begin{eqnarray}
&&\log_{10}\left(\frac{M_{\rm rem}}{M_\odot}\right) = \nonumber \\
&&\left\{ \begin{array}{ll}
\log_{10}(0.03357x_{\rm CO} + 1.31780) & (x_{\rm CO}<6.357),\\
-0.02466x_{\rm CO}+1.28070 & (6.357<x_{\rm CO}<7.311), \\
\log_{10}(0.03357x_{\rm CO} + 1.31780) & (7.311<x_{\rm CO}<12.925),
\end{array}\right.
\label{eq:remnant_mass}
\end{eqnarray}
where $x_{\rm CO}=M_{\rm CO}/M_\odot$. The intermediate range is actually expected to form black holes as remnants, possibly with much weaker explosion energies as low as $E_{\rm ej}=10^{46}$--$10^{48}$ ergs and smaller ejecta mass \citep{Lovegrove13,Fernandez18,Tsuna20,Ivanov21}. The code is nonetheless only capable with values of $E_{\rm ej}$ around $10^{51}$ ergs seen in canonical supernovae (see Table \ref{table:Parameters}), because our assumption of homologous ejecta ($\upsilon=r/t$) breaks down for such low-energy explosions. This is especially the case for red supergiants, which takes months for the ejecta to expand to twice the progenitor's radius.

The inner exponent $\delta$ typically takes a value between $0$ and $1$ \citep{Matzner99}, and here $\delta$ is fixed to be $1$. The value of the outer exponent $n$ is determined by first obtaining a ``global" polytropic index $N_{\rm pol}$ inside the envelope at core-collapse that best fits the relation $p=K\rho^{1+1/N_{\rm pol}}$, where constants $K$ and $N_{\rm pol}$ are the fitting parameters. Then we use a formula connecting $N_{\rm pol}$ and $n$ as (\citealt{Matzner99}; equation 25)
\begin{eqnarray}
 n = \frac{N_{\rm pol}+1+\beta N_{\rm pol}}{\beta N_{\rm pol}},\label{eq:npol_n}
\end{eqnarray}
where $\beta\approx 0.19$ only weakly depends on $N_{\rm pol}$.

The abundance is assumed to be uniform throughout the computational region, and the code takes the representative helium mass fraction $Y$ as the average of that inside the CSM. We fix the metal mass fraction $Z$, and adopt the solar abundance of $Z_{\odot}=0.014$ \citep{2009ARA&A..47..481A}. The hydrogen mass fraction is then obtained as $X=1-Y-Z$.

\subsubsection{Shock Structure}
\label{sec:lightcurve_shock}
The CHIPS code calculates the bolometric and multi-band LCs at an observer in two steps. The code first calculates the shock structure as a function of radius $r$ for each of the shocks, by the following equations at the shock rest frame that assumes steady state.
\begin{eqnarray}
 &&\frac{\partial (r^2\rho \upsilon)}{\partial r} = 0, \label{eq:cont}\\
 &&\upsilon\frac{\partial \upsilon}{\partial r} + \frac{1}{\rho}\frac{\partial p}{\partial r} = 0, \label{eq:mom}\\
 &&\frac{\partial}{\partial r} \left[r^2\left\{\rho \upsilon\left(\frac{1}{2}\upsilon^2+e+\frac{p}{\rho}\right)+F\right\}\right] = 0.\label{eq:ene}
\end{eqnarray}
For simplicity we assume gas and radiation are in thermal equilibrium, and set the pressure $p$ and specific internal energy $e$ as a function of $\rho$ and temperature $T$ ($=T_\mathrm{g}=T_\mathrm{r}$) as
\begin{eqnarray}
 p &=& \frac{\rho}{\mu m_u}k_BT+\frac{1}{3}aT^4 \\
 \rho e &=& \frac{3}{2}\frac{\rho}{\mu m_u}k_BT + aT^4,
\end{eqnarray}
where $\mu,\, m_u,\, k_B,\, a,\,T_\mathrm{g},\,T_\mathrm{r}$ are respectively the mean molecular weight, the atomic mass unit, the Boltzmann constant, the radiation constant, the gas temperature, and the radiation temperature.

Though the radiative flux $F$ was calculated by assuming a diffusion approximation in \cite{Takei20}, we refine this and adopt a flux-limited diffusion \citep{Levermore81}
\begin{eqnarray}
 F &=& -\frac{c}{\kappa_{R}\rho}\lambda \frac{\partial E}{\partial r}, \label{eq:flux_limiter}\\
 \lambda &=& \frac{2+R}{6+3R+R^2}, R = \frac{|\partial E/\partial r|}{\kappa_{R}\rho E},
\end{eqnarray}
to extend our model to the optically thin regime, where $E=aT^4$, $c$ is the speed of light, and $\kappa_{R}$ is the Rosseland mean opacity.
%For the absorption opacity $\kappa_{R}$, we adopt the Rosseland mean opacity and use the TOPS opacity table \citep{1995ASPC...78...51M}, which tabulates $\kappa$ as a function of density and temperature. For   {the scattering opacity} $\sigma$, we assume that Thomson scattering is dominant and set $\sigma= {0.2\ {\rm cm^2\ g^{-1}}}(1+X)$.
For the Rosseland mean opacity $\kappa_{R}$, we use the TOPS opacity table \citep{1995ASPC...78...51M}, which tabulates $\kappa_{R}$ as a function of density and temperature.

To solve the hydrodynamical equations (\ref{eq:cont})-(\ref{eq:ene}), we first set boundary conditions at the immediate downstream of each shock by the jump conditions, which can be written in the rest frame of the shock as
\begin{eqnarray}
\rho_{\rm down}\upsilon_{\rm down} &=& \rho_{\rm up}\upsilon_{\rm up}, \\ 
\rho_{\rm down}\upsilon_{\rm down}^2+p_{\rm down} &=& \rho_{\rm up}\upsilon_{\rm up}^2+p_{\rm up}, \\
\frac{1}{2}\upsilon_{\rm down}^2+e_{\rm down}+\frac{p_{\rm down}}{\rho_{\rm down}} &=& \frac{1}{2}\upsilon_{\rm up}^2+e_{\rm up}+\frac{p_{\rm up}}{\rho_{\rm up}},
\end{eqnarray}
where the subscripts down (up) denote variables at the downstream (upstream) of the shock. The upstream density $\rho_{\rm up}$ for the forward (reverse) shock is given from the density profiles in the CSM (ejecta), and $p_{\rm up}, e_{\rm up}$ are assumed to be $0$ for both shocks.

There are three unknown parameters in the equations: the velocities of the two shocks and the radiative flux at the forward shock.
% {The flux at the reverse shock is significantly small compared to that at the forward shock in the early phase since the shocked region is optically thick \citep{Tsuna19}.}
 {The radiative flux at the reverse shock front is set to 0 since the reverse shock weakly dissipates the kinetic energy into radiation compared to the forward shock.}
We derive these three values so that the velocity, pressure,  and flux become continuous at the contact discontinuity by integrating the equations iteratively. The detailed procedure to calculate the temporal evolution of the shock waves is given in \citet{Takei20}. %The location of the contact discontinuity is independently obtained by equating the mass of the shocked ejecta and the total mass of the ejecta that has been swept up by the reverse shock $\int_{r_{\rm rs}}^\infty 4\pi r^2 \rho_{\rm ej}(r,t)dr$. The obtained shock velocities are used to obtain the positions of the shocks at the next timestep.

One of the most important updates from \cite{Takei20} is on the initial condition. In order to start the calculation, we need to set the initial position of the reverse shock $r_0$ and initial time $t_0$. \citet{Takei20} fixed $t_0=1$ day for simplicity. The assumption of a steady-state shock can be violated at early phases when the photon diffusion is slower than the expansion of the shocked shell, which we have seen to occur for a large $f_{\rm inj}$ or small $t_{\rm inj}$. %valid only when the dynamical timescale is longer than the cooling and diffusion timescales (i.e. the features of radiation emitted from the shock at time $t$ only weakly depends on the dynamics before $t$). This assumption is 
We will see the situation where the photon diffusion time becomes comparable to the expansion time scale of the shocked shell by using a simple model and explain a method to deal with this situation below.  

%The radius where the diffusion time and shock expansion time become comparable can be estimated as follows. 
As seen in equation (\ref{eq:rho_CSM}), the inner density profile of the CSM closely follows $\rho\propto r^{-1.5}$. Initially the shocked region is well approximated to be adiabatic with an index $\gamma=4/3$, and a self-similar solution for the shock propagation in a CSM with $\rho=qr^{-1.5}$ \citep{Chevalier82} gives the radius and the velocity of the shock front as functions of time as 
\begin{eqnarray}
    r_{\rm sh} &=& \left(\frac{Ag^n}{q}\right)^{1/(n-1.5)}t^{(n-3)/(n-1.5)} \label{eq:r_sh_Chevalier}\\
    \upsilon_{\rm sh} &=& \frac{dr_{\rm sh}}{dt} = \frac{n-3}{n-1.5}\frac{r_{\rm sh}}{t},
\end{eqnarray}
where $A$ is a constant dependent on $n$ and $\gamma$. The value of $q$ is determined by the total mass $M_{\rm CSM}$ of the CSM distributed within a radius $r_{\rm out}$ as
\begin{equation}
q\approx 7.5\times 10^8\ {\rm cgs}\left(\frac{M_{\rm CSM}}{10^{-1}M_\odot}\right) \left(\frac{r_{\rm out}}{10^{15}\ {\rm cm}}\right)^{-1.5}.
\label{eq:q_vs_M_t}
\end{equation}
The optical depth of the unshocked CSM is given as
\begin{eqnarray}
 \tau_{\rm CSM}(r_{\rm sh})=\int^\infty_{r_{\rm sh}} \bar\kappa\rho dr \approx 2\bar\kappa q\left(r_{\rm sh}^{-0.5}-r_*^{-0.5}\right)
\end{eqnarray}
where the average opacity $\bar\kappa$ is assumed to be given by the Thomson opacity. The latter equation assumes that the profile $\rho\propto r^{-1.5}$ continues up to a certain radius $r_*>r_{\rm sh}$, and that the contribution beyond $r_*$ is negligible\footnote{In the CHIPS code the exact integrated value of $\tau_{\rm CSM}$ is calculated, but to gain insight we adopt this approximation in the remaining part of this section.}.

\begin{figure}
 \centering
 \includegraphics[width=\linewidth]{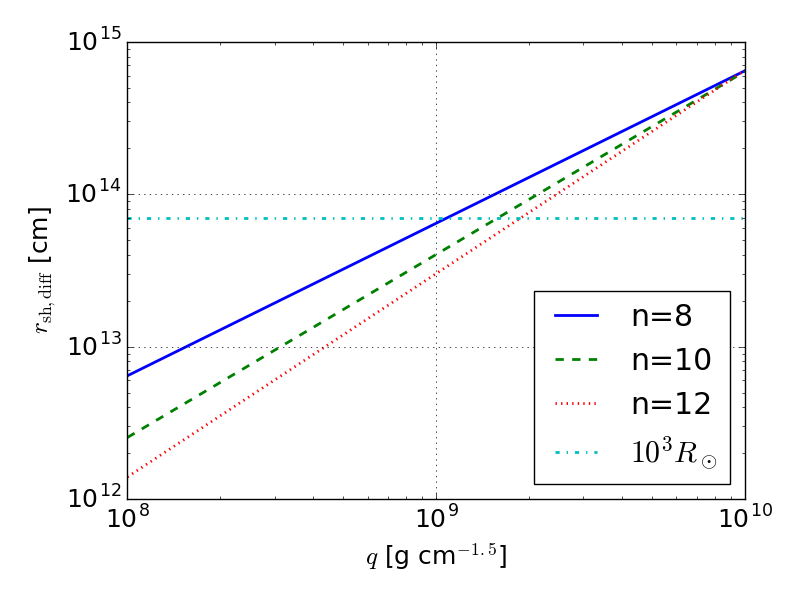}
\caption{Dependence of $r_{\rm sh,\,diff}$ as a function of the CSM density normalization parameter $q$. Here the typical values $E_{\rm ej}=10^{51}$ erg, $M_{\rm ej}=10\ M_\odot$, $\delta=1$, $\bar\kappa=0.34\ {\rm cm^2\ g^{-1}}$ were assumed. Plotted as dash-dotted horizontal line is $r=10^3\ R_\odot$, a typical value of $2R_*$ for red supergiants.}
 \label{fig:rshdiff}
 \end{figure}

We further assume $r_*\gg r_{\rm sh}$ so that the second term in the parenthesis can be neglected. Therefore the diffusion and dynamical timescales become comparable when $t$ and $r_{\rm sh}$ satisfy the following condition
%\begin{eqnarray}
%    \frac{c}{\tau_{\rm CSM}}=\frac{n-3}{n-1.5}\frac{r_{\rm sh}}{t} \iff t=\frac{2(n-3)}{3(n-1.5)}\frac{\kappa q r_{\rm sh}^{0.5}}{c}.
%\end{eqnarray}
\begin{eqnarray}
 \frac{\tau_{\rm CSM}r_{\rm sh}}{c} = \frac{r_{\rm sh}}{\upsilon_{\rm sh}} \iff t=\frac{2(n-3)}{3(n-1.5)}\frac{\bar\kappa q r_{\rm sh}^{0.5}}{c}.
\end{eqnarray}
Substituting this into equation (\ref{eq:r_sh_Chevalier}) we solve for $r_{\rm sh}$ and obtain
\begin{eqnarray}
    r_{\rm sh,\,diff}= \left(Ag^nq^{n-4}\right)^{2/n}\left[\frac{2(n-3)\bar\kappa}{3(n-1.5)c}\right]^{2(n-3)/n}.
    \label{eq:r_sh_diff}
\end{eqnarray}
We plot $r_{\rm sh\,diff}$ as a function of $q$ in Figure \ref{fig:rshdiff} for three values of $n$ that are realistic for stellar progenitors. We see that the steady-state assumption is invalid at $r=2R_*$ for $q\gtrsim 10^9$ cgs. If the CSM expands for a time $t_{\rm inj}$ with a velocity comparable to the escape velocity at the stellar surface $\upsilon_\mathrm{w}\sim\sqrt{2GM_*/R_*}$, then the CSM will extend out to $r_{\rm out}\sim \upsilon_\mathrm{w}t_{\rm inj}$. Thus the radiative equilibrium is not realized at around the forward shock in the CSM with the corresponding mass
\begin{eqnarray}
 M_{\rm CSM} \gtrsim 0.1\ M_\odot \left(\frac{t_{\rm inj}}{3\ {\rm yr}}\right)^{1.5}  \left(\frac{\upsilon_\mathrm{w}}{100\ {\rm km \ s^{-1}}}\right)^{1.5},
\end{eqnarray}
 until the forward shock reaches the radius $r_{\rm sh,\,diff}$.
We thus set $r_0$ to be the larger of $2R_{*}$ and $r_{\rm sh,\,diff}$. Then following \cite{Takei20} we adopt an analytical model of shock propagation \citep{Moriya_et_al_13} to estimate the corresponding $t_0$.

In the case of $r_{\rm sh,\,diff}>2R_*$, the shock breaks out at the radius $r_{\rm sh, diff}$ inside the CSM. The internal energy $E_{\rm int,\,bo}$ that has been stored as radiation in the shocked region before the shock breakout is released on roughly the dynamical time. To take this effect into account, we inject this internal energy as a source term at the inner boundary ($r=r_{\rm fs}$) when solving the radiative transfer equation in the unshocked CSM outlined in the next section. The energy is injected depending on time in the form of an energy flux as \citep{Arnett80,Smith07}
\begin{eqnarray}
F_{\rm bo}(t)\propto\frac{E_{\rm int,\,bo}}{4\pi r_{\rm sh}(t)^2t_0}\exp\left[-\frac{t}{t_0}-\frac{1}{2}\left(\frac{t}{t_0}\right)^2\right].
\label{eq:breakout_flux}
\end{eqnarray}
The internal energy stored in the shocked region at time $t$ is proportional to the ejecta's kinetic energy dissipated up to time $t$, and the proportionality factor depends on the CSM density profile within the breakout radius, $\gamma$, and $n$. We fix the first two as $\rho\propto r^{-1.5}$ and $\gamma=4/3$, and tabulate the value of this factor for various values of $n$ using the self-similar solution \citep{Chevalier82}.
For a given value of $n$ we obtain the energy $E_{\rm int,\,bo}$ through interpolation.

\subsubsection{Radiative Transfer in Unshocked CSM}
By solving equations for the shock structure at each time step, we obtain the flux $F$ at the forward shock as a function of time from equation (\ref{eq:flux_limiter}).
To obtain the thermal structure inside the unshocked CSM with this flux and the LCs observed by a distant observer, the CHIPS code solves the following two-temperature radiative transfer and energy equations for $E(=aT_\mathrm{r}^4)$, and $U\left(=3k_{B}T_\mathrm{g}/(2\mu m_\mathrm{u})\right)$,
\begin{eqnarray}
&& \frac{\partial E}{\partial t}+\frac{\partial(r^{2}F)}{r^{2}\partial r}=\kappa_\mathrm{a}\rho c(aT_\mathrm{g}^{4}-E),\label{eqn:radtra}\\
&& \left(\frac{\partial}{\partial t}+\upsilon\frac{\partial}{\partial r}\right)U+p_\mathrm{g}\upsilon\frac{\partial\rho^{-1}}{\partial r}=\kappa_\mathrm{a}c(E-aT_\mathrm{g}^{4}),\label{eqn:energy}
\end{eqnarray}
where $U$ denotes the specific internal energy of the gas, $p_\mathrm{g}$ the gas pressure, and $\kappa_\mathrm{a}$ the absorption opacity. 
%The flux $F$ is obtained from equation (\ref{eq:flux_limiter}).

The boundary condition for the radiative flux is obtained as follows. Integrating equation (\ref{eqn:radtra}) over the volume enclosed between $r=r_\mathrm{fs}$ and $r=r_\mathrm{out}$ yields
\begin{eqnarray}
\int_{r_\mathrm{fs}}^{r_\mathrm{out}}\frac{\partial(4\pi r^{2}E)}{\partial t}dr\simeq4\pi r_\mathrm{fs}^{2}F(r_\mathrm{fs})-4\pi r_\mathrm{out}^{2}F(r_\mathrm{out}),
\end{eqnarray}
where the source term is omitted.
From this equation, we obtain the time derivative of the total radiation energy in this volume as 
\begin{eqnarray}
&&\frac{\partial}{\partial t}\int_{r_\mathrm{fs}}^{r_\mathrm{out}}4\pi r^{2}Edr=\int_{r_\mathrm{fs}}^{r_\mathrm{out}}\frac{\partial(4\pi r^{2}E)}{\partial t}dr-4\pi r_\mathrm{fs}^2E(r_\mathrm{fs})u_\mathrm{fs} \nonumber \\
&&\simeq -4\pi r_\mathrm{out}^{2}F(r_\mathrm{out})+4\pi r_\mathrm{fs}^{2}(F(r_\mathrm{fs})-E(r_\mathrm{fs})u_\mathrm{fs}),
\end{eqnarray}
Since the last term $F(r_\mathrm{fs})-E(r_\mathrm{fs})u_\mathrm{fs}$ denotes the flux at the shock front, this should be equal to $F$ at the forward shock (Otherwise the radiation energy would be lost or generated at the shock). Thus we obtain
\begin{equation}
    F(r_\mathrm{fs})=F+E(r_\mathrm{fs})u_\mathrm{fs},
\end{equation}
as the boundary condition.
%since the radiative transfer in the unshocked CSM is done by post-process.
%In order to compensate this, we add this energy flux term to $F(r_\mathrm{fs})$.
% {where $u_\mathrm{fs}$ denotes the velocity of the forward shock front. Since $F(r_\mathrm{fs})$ is the value in the rest frame of the forward shock front, we convert this to that in the rest frame of the center of the coordinate system by adding $E(r_\mathrm{fs})u_\mathrm{fs}$.}

\citet{Takei20} calculated the absorption opacity $\kappa_\mathrm{a}$ by subtracting the scattering opacity from the OPAL Rosseland mean opacity \citep[OPAL opacities:][]{Ross}.
However, the above opacity is not realistic as the absorption opacity for the source term in equations (\ref{eqn:radtra}), (\ref{eqn:energy}) since the Rosseland mean opacity is derived assuming that the region is optically thick and used for the diffusion approximation.
Therefore, the Planck mean opacity is newly implemented in CHIPS as the absorption opacity from the TOPS opacity table \citep{1995ASPC...78...51M} in order to more precisely calculate the source term.
We generated the Rosseland and Planck mean opacities for helium fractions of $Y=0,\,0.1,\,\cdots,\,0.7$ with a fixed metallicity of $Z=Z_{\odot}$.
For other helium abundances, opacities are generated by linear interpolation.

We also implement multigroup radiative transfer calculation using post-process ray-tracing method to obtain multi-color LCs at $U,\,B,\,V,\,R,\,I$ bands. For ray-tracing we follow the methods in \citet{2015PASJ...67...54K} and \citet{Suzuki21}. We first integrate the radiative transfer equation along each ray passing through the CSM with a different impact parameter $b$,
\begin{eqnarray}
\frac{dI_{\nu}}{ds}=(\kappa_{\nu}+\sigma_{\nu})\rho(S_{\nu}-I_{\nu}),
\end{eqnarray}
where $I_{\nu},\,S_{\nu}$ are the intensity and the source function.
$\kappa_{\nu},\,\sigma_{\nu}$ are the absorption and scattering opacity at frequency $\nu$.
We can integrate this equation with respect to the optical depth $d\tau_{\nu}=(\kappa_{\nu}+\sigma_{\nu})\rho ds$ and obtain the formal solution,
\begin{eqnarray}
I_{\nu}(b,\tau_{\nu})=I_{\nu}(b,0)e^{-\tau_{\nu}}+\int_{0}^{\tau_{\nu}}B_{\nu}(T_\mathrm{g})e^{-(\tau_{\nu}-\tau'_{\nu})}d\tau'_{\nu},\label{eqn:formal_solution}
\end{eqnarray}
where we assume that $S_{\nu}$ is the Planck function $B_{\nu}(T_\mathrm{g})$.
\begin{figure}
\centering
\includegraphics[width=0.8\linewidth]{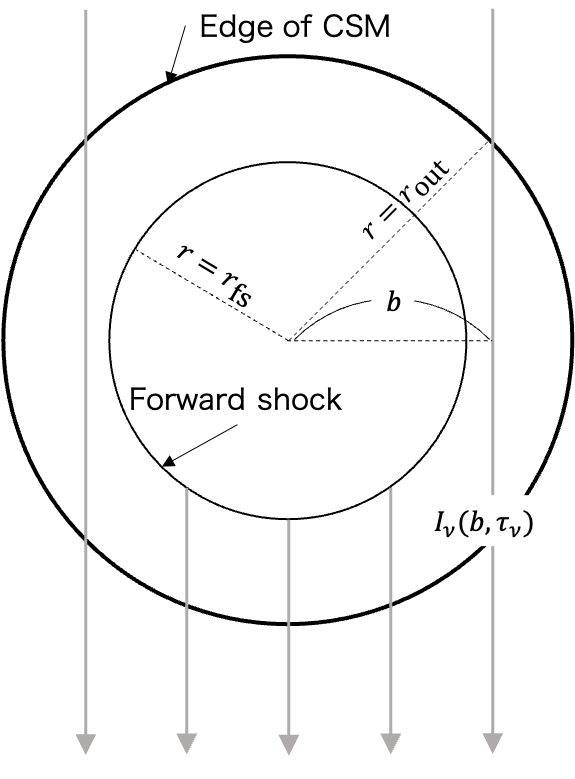}
\caption{Schematic view of the ray-tracing method used to obtain the multi-band light curve. To obtain the intensity $I_{\rm \nu}(b,\tau_{\nu})$ at each impact parameter $b$, we solve the radiation transfer equation along rays crossing through the CSM for $b>r_{\rm fs}$, and rays from $r=r_{\rm fs}$ for $0<b<r_{\rm fs}$. The rays are shown as gray arrows.}
\label{fig:ray_tracing}
\end{figure}
Again, we use the TOPS opacity for the frequency-dependent absorption opacity to keep the consistency with the Planck mean opacity. We obtain $I_\nu(b,0)$ from the flux emerging from the forward shock by assuming that the spectral shape is blackbody for simplicity if a ray penetrates the shocked region.
On the other hand, along rays with $b>r_\mathrm{fs}$, the boundary value of $I_\nu$ is 0 since the ray emanates from infinity.
We can summarize the boundary conditions for $I_{\nu}$ as
\begin{eqnarray}
I_{\nu}(b,0)=\left\{ \begin{array}{ll}
0 & (b>r_\mathrm{fs}), \\
B_{\nu}(T_\mathrm{fs}) & (0\leq b \leq r_\mathrm{fs}),
\end{array}
\right.
\end{eqnarray}
where $T_\mathrm{fs}$ satisfies $F(r_\mathrm{fs})=(ac/4)T_\mathrm{fs}^{4}$.
Then, equation (\ref{eqn:formal_solution}) is integrated from $\tau_{\nu}=0$. We note that the exact spectral shape should depend on the density of the CSM. A less dense CSM {, for instance, that of a stellar wind with a mass-loss rate of $\lesssim 10^{-2}M_\odot\,{\rm yr^{-1}}$ for a wind velocity of $\upsilon_{w}=100\,{\rm km\,s^{-1}}$,} is expected to produce a harder spectrum that extends to X-rays \citep{Chevalier12,Svirski12,Tsuna21}.

We calculate the luminosity per unit frequency, $L_{\nu}$ by integrating the surface intensity $I_\nu(b)=I_\nu(b,\tau_\nu(b))$ at each impact parameter $b$ over the surface as,
\begin{eqnarray}
L_{\nu}=8\pi^{2}\int_{0}^{b_\mathrm{max}}I_{\nu}(b)bdb,
\end{eqnarray}
where $b_\mathrm{max}=r_\mathrm{out}$.
Finally we can calculate the absolute magnitude at each band, using a filter function in \citet{filter_function}.

\section{CHIPS Results}
\label{sec:Results}
Here we present some results from a series of simulations performed with the CHIPS code and explore the dependence of LCs of interaction-powered SNe on four key parameters: the mass $M_*$ of the progenitor, the injected energy $f_\mathrm{inj}$ for the precursor mass eruption and its timing $t_\mathrm{inj}$, and the explosion energy $E_\mathrm{ej}$ of the SN. First, we present and discuss results of mass eruptions with various parameter sets. Then we present resultant LCs to summarize the dependence of some observable quantities on the key parameters, which will be useful to discuss the origin of an observed interaction-powered transient. We compare CHIPS results with other works in Appendix \ref{sec:cmp_with_snec}.

\begin{figure*}
 \centering
 \includegraphics[width=\linewidth]{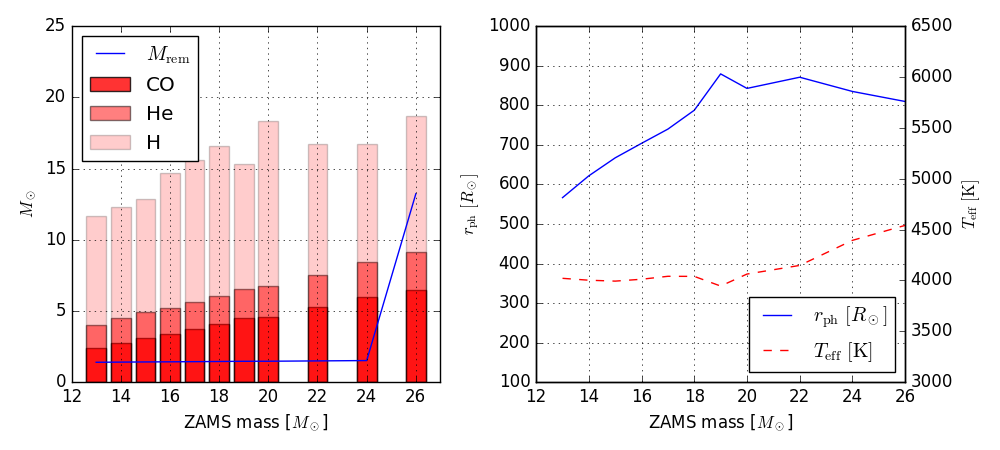}
\caption{Properties of the progenitors we adopt in this work. The left panel shows the masses of the hydrogen-rich envelope, helium core and the carbon-oxygen core, and the remnant mass $M_{\rm rem}$ obtained from equation (\ref{eq:remnant_mass}). The right panel shows photospheric radius (in $R_\odot$) of each star and the effective temperature (in Kelvins). These progenitors are available when downloading the CHIPS code, and one can use them without running the MESA calculation. }
 \label{fig:progenitors}
 \end{figure*}

In what follows, we present our results for massive stars evolved with the \verb|example_make_pre_ccsn| test suite in MESA version 12778. To reduce the dimensionality of our parameter space, we have restricted the metallicity to be solar and neglected stellar rotation. The properties of the progenitors that we have created are summarized in Figure \ref{fig:progenitors}.

\subsection{Mass Eruption: Dependence on the Progenitor and Injected Energy}
\begin{figure}
\centering
\includegraphics[width=\linewidth]{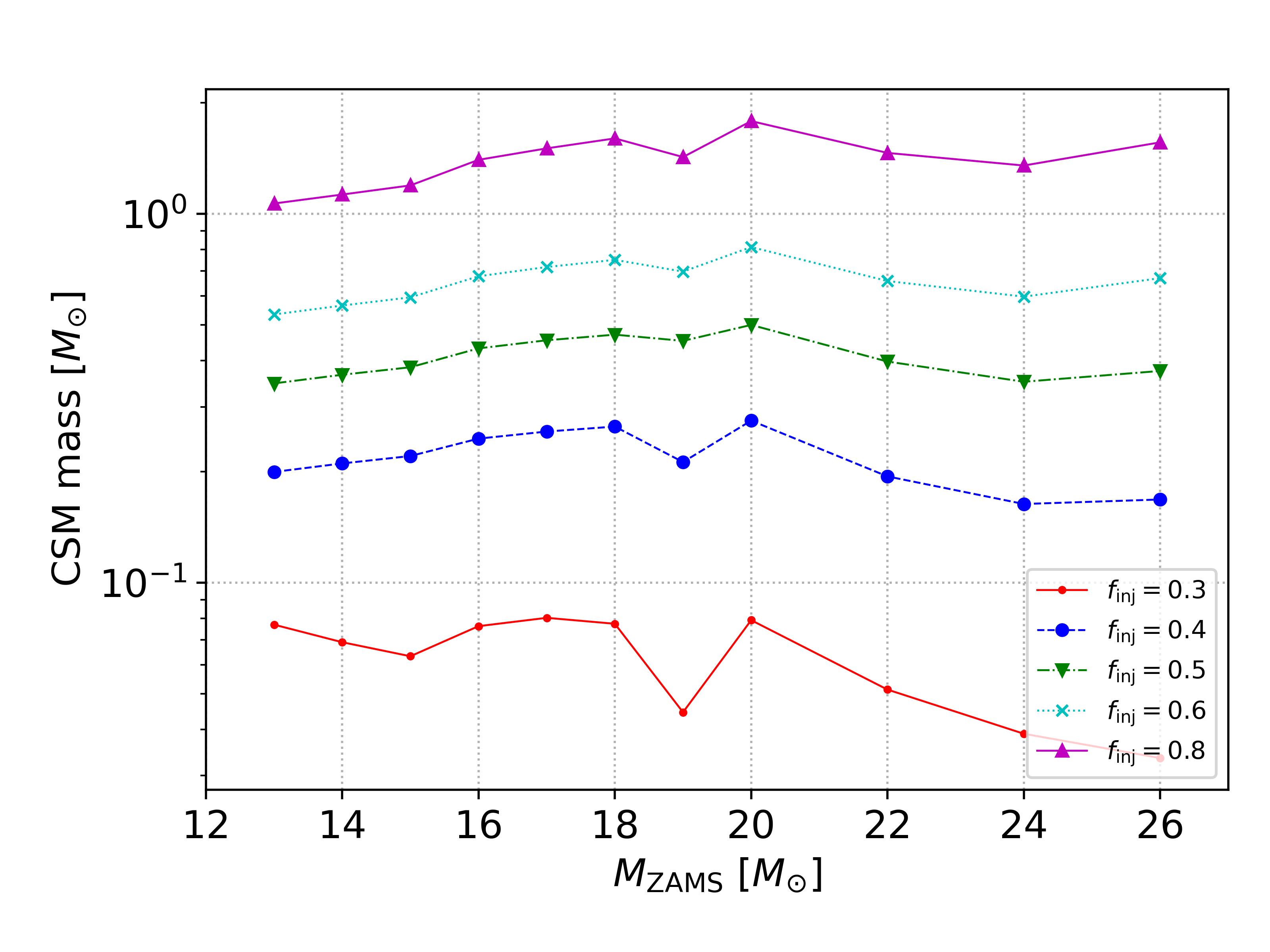}
\caption{The erupted mass as a function of $f_\mathrm{inj}$ at $t_\mathrm{inj}=10\,{\rm yrs}$. The CSM mass decreases with increasing $t_\mathrm{inj}$ due to the fallback of the bound mass elements towards the progenitor.}
\label{fig:erupted_mass}
\end{figure}
\begin{figure}
\centering
\includegraphics[width=\linewidth]{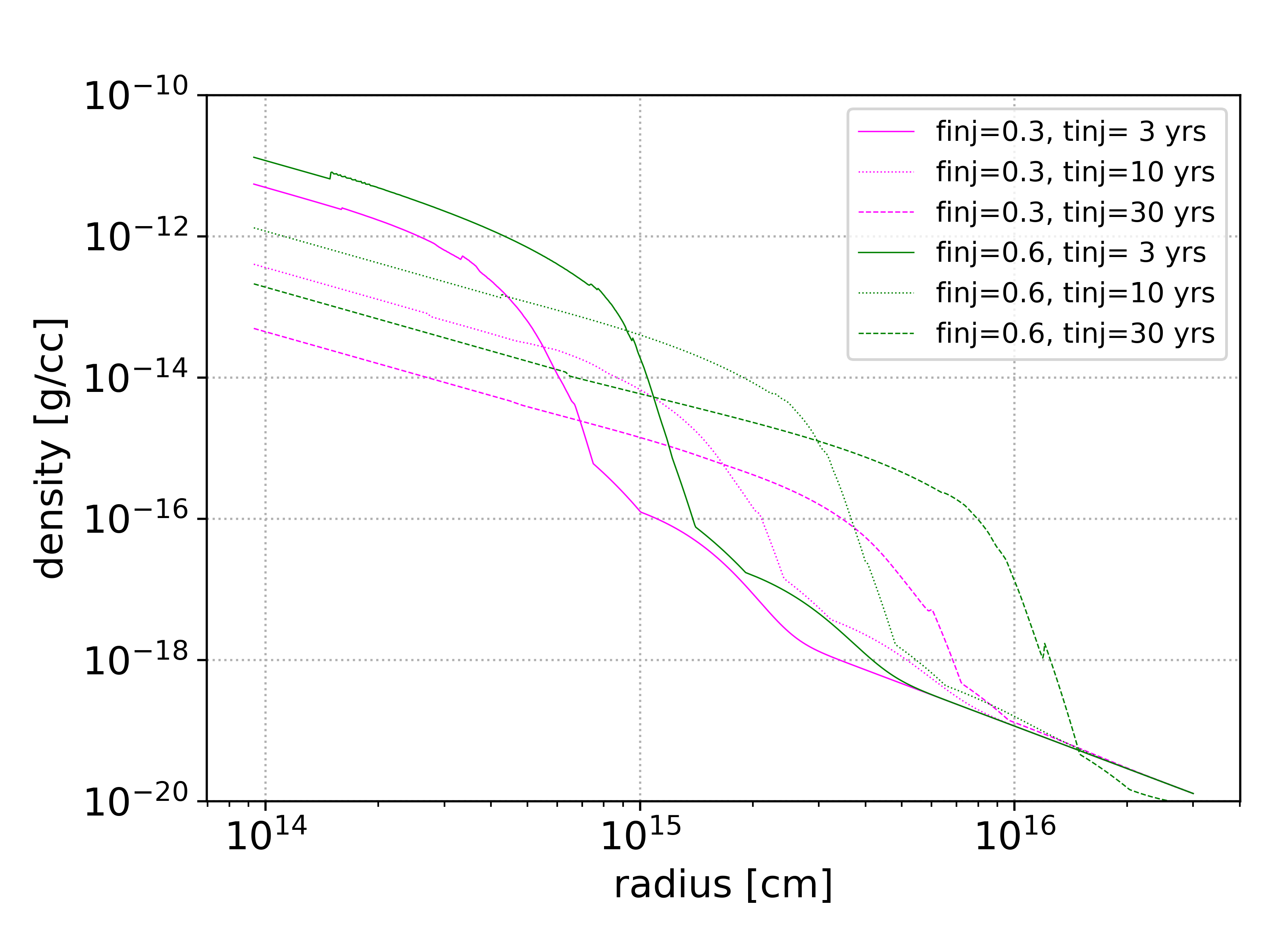}
\caption{Comparison of CSM density profiles as functions of radius for different $f_\mathrm{inj}$ and $t_\mathrm{inj}$. The steady wind is smoothly attached to the outer edge of the erupted material.}
\label{fig:CSM_cmp}
\end{figure}
Figure \ref{fig:erupted_mass} shows the dependence of erupted mass on the injected energy $f_\mathrm{inj}$. 
As shown in this figure, the erupted mass increases with $f_\mathrm{inj}$ for all the models we create.
The erupted mass of model $M_{*}=19M_\odot$ is lower than those of models $M_{*}=17,\,18M_\odot$.
The photospheric radius plotted in Figure \ref{fig:progenitors} is large at $M_{*}=19M_\odot$, which means that the shock breakout occurs at more inner part of the envelope compared to other progenitor models.
Due to the inefficient energy transport towards the outer envelope, the erupted mass becomes small.
This tendency is prominent for lower $f_\mathrm{inj}$.
%This is caused by the fallback towards the progenitor due to the strong gravity compared to the other models.
%This tendency is prominent for lower $f_\mathrm{inj}$ because the velocity of the most of mass elements has not reached the escape velocity.

The density profiles of the remeshed CSM for some parameters $f_\mathrm{inj},\,t_\mathrm{inj}$ are plotted in Figure \ref{fig:CSM_cmp}.
While the density profile at inner region well follows $\rho\propto r^{-1.5}$ at $t_\mathrm{inj}=10,\,30\,{\rm yrs}$, the exponent slightly deviates from $-1.5$ at $t_\mathrm{inj}=3\,{\rm yrs}$. This is because the gravity of the progenitor does not control the motion of the inner region yet or the matter does not start to fall back yet. It takes significantly longer time than the free-fall time scale given by $\sim\sqrt{r^3/GM_*}\sim0.9\,\mathrm{yr}(r/10^{14}\,\mathrm{cm})^{3/2}(M_*/10\,M_\odot)^{-1/2}$ for the inner region to enter this fall-back phase \citep{Tsuna21b}.
%Substituting $t_\mathrm{inj}=3\,{\rm yrs}$ into $X$ defined in \citet{Tsuna21b}, we obtain $X\sim0.2$.
%Since this power-law holds for where $X\ll 1$, the profile has not followed $\rho\propto r^{-1.5}$ yet at $t_\mathrm{inj}=3\,{\rm yrs}$.

\subsection{Supernova Light Curves}
\begin{figure}
\centering
\includegraphics[width=\linewidth]{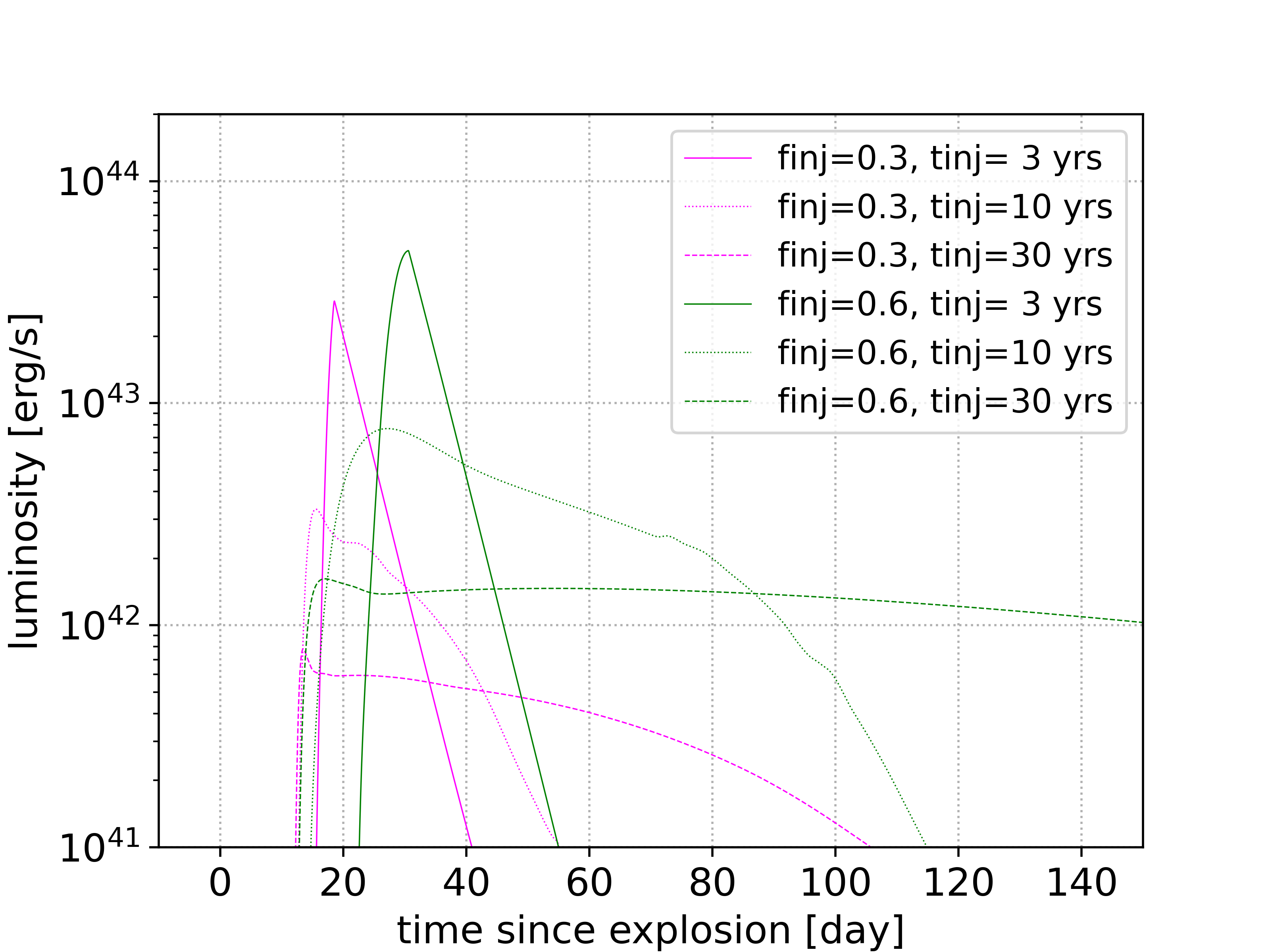}
\caption{Sample light curves for the case $M_*=15M_{\odot},\,E_{\rm ej}=10^{51}\,{\rm erg}$ when $f_{\rm inj}$ and $t_{\rm inj}$ are varied as indicated in the panel.}
\label{fig:15Msun_lc_51erg}
\end{figure}

%\subsubsection{Bolometric light curves}
Bolometric LCs are a useful probe of SNe because i) bolometric LCs can be obtained from a simplified numerical model and ii) the bolometric LC constructed from photometric observations include important information about SN events such as the heating source, the size of the progenitor, the explosion energy, the CSM structure, etc. The CHIPS code provides bolometric LCs for interaction-powered SNe, as shown for several example sets of parameters in Figure \ref{fig:15Msun_lc_51erg}. We focus on two observables that are often used to characterize the LC: peak luminosity $L_{\rm peak}$ and rise time $t_{\rm rise}$. In this subsection we summarize the dependence of these observables on the input parameters of CHIPS.
\subsubsection{Dependence on the Timing of Eruption $t_{\rm inj}$}\label{sec:tinj}
\begin{figure*}
\centering
\includegraphics[width=\linewidth]{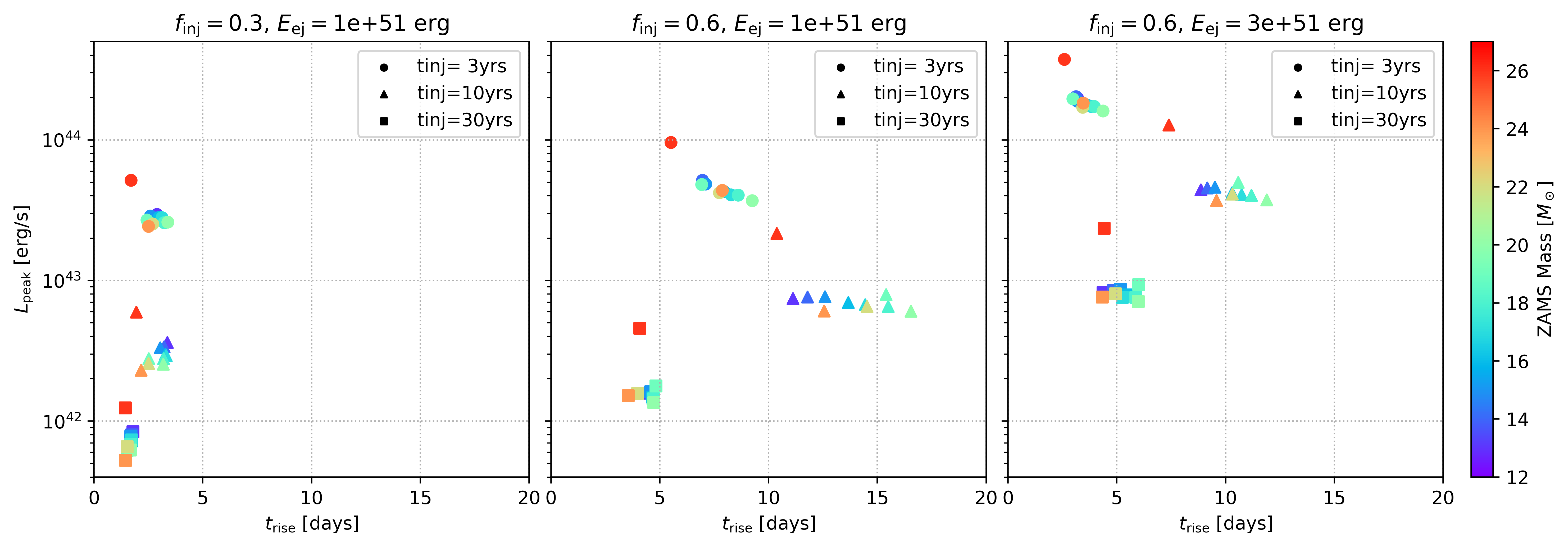}
\caption{The dependence of $t_\mathrm{rise}$--$L_\mathrm{peak}$ relation on $t_\mathrm{inj}$. The other parameters selected are shown on the top of each panel.}
\label{fig:tinj_cmp}
\end{figure*}
Figure \ref{fig:tinj_cmp} shows the dependence of $L_\mathrm{peak}$ and $t_\mathrm{rise}$ on $t_\mathrm{inj}$. We have defined $t_\mathrm{rise}$ to be the elapsed time from the luminosity being $0.01L_{\rm peak}$ to $L_{\rm peak}$.
As seen from the figure, $L_\mathrm{peak}$ becomes larger for shorter $t_\mathrm{inj}$, while the dependence of $t_{\rm rise}$ on $t_\mathrm{inj}$ is more complex.

The rise time depends on the transition radius $r_{*}$ and the shock breakout radius $R_{d}\equiv (2\bar\kappa q\upsilon_\mathrm{fs}/c)^{2}$, which is equal to $r_{\rm sh, diff}$ in the limit of large $n$. For convenience we define a ratio of these two radii
\begin{eqnarray}
% 41.6
\frac{R_{d}}{r_{*}}&\approx&42\left(\frac{\bar\kappa}{0.34\,{\rm cm^{2}\,g^{-1}}}\right)^{2}\left(\frac{q}{10^{10}\,{\rm g\,cm^{-1.5}}}\right)^{2} \nonumber \\
&\times&\left(\frac{\upsilon_\mathrm{fs}/c}{0.03}\right)^{2}\left(\frac{r_{*}}{10^{15}\,{\rm cm}}\right)^{-1},\label{eq:R_d}
\end{eqnarray}
where we have assumed for simplicity that the CSM extends as a power law $\rho\propto r^{-1.5}$ and truncates at $r=r_*$. Following the calculations of \cite{Chevalier_Irwin_2011}, we can analytically obtain the rise time for the profile $\rho\propto r^{-1.5}$ as
\begin{eqnarray}
t_{\rm rise}
&\approx& \frac{R_d}{\upsilon_{\rm fs}}\times \left\{ \begin{array}{ll}
1 & (R_{d}/r_{*} < 1),\\
2(R_d/r_*)^{-1.5} & (R_{d}/r_{*} > 1).
\end{array}\right.
\label{eq:ana_rise_time}
\end{eqnarray}
As these two are discontinuous, we instead adopt their harmonic mean
\begin{eqnarray}
\tilde t_{\rm rise} \equiv \frac{R_d}{\upsilon_{\rm fs}}\left[\frac{1}{1+0.5(R_d/r_*)^{1.5}}\right],
\end{eqnarray}
which peaks at $R_d/r_*\approx 1$ when $r_*$ is fixed.

To see if the numerical results from CHIPS match this expectation, we test two cases with different $t_{\rm inj}$ of 10 years and 3 years, and the other parameters fixed as $M_*=15M_\odot,\,f_\mathrm{inj}=0.8,\,E_\mathrm{ej}=10^{51}\,{\rm erg}$. For $t_{\rm inj}=10$ years, we obtain from the analytical model $R_{d}/r_{*}\approx 0.21<1$ and $\tilde t_\mathrm{rise}\approx16\,{\rm days}$. The numerical results from CHIPS show a good agreement with this, with $t_\mathrm{rise}\approx16\,{\rm days}$.
On the other hand, changing $t_{\rm inj}$ to 3 years gives $R_{d}/r_{*}=80\gg 1$ and $\tilde t_\mathrm{rise}\approx 8.6\,{\rm days}$. The numerical simulation nicely matches as well in this case, with $t_\mathrm{rise}\approx 9.0\,{\rm days}$.

The analytical model in equation (\ref{eq:ana_rise_time}) implies that for an optically thick CSM the rise time peaks at a certain $t_{\rm inj}$ for fixed $f_\mathrm{inj},\,M_*,\,E_\mathrm{ej}$. We can derive the relation $\rho_{*}\propto t_\mathrm{inj}^{-3},\,r_{*}\propto t_\mathrm{inj}$ since around $r\sim r_{*}$ the CSM expands homologously\footnote{The exact dependence of $\rho_*$ on $t_{\rm inj}$ slightly deviates from $\rho_{*}\propto t_\mathrm{inj}^{-3}$ (see Appendix A of \citealt{Tsuna21b}), but this difference is unlikely to significantly affect the discussion.}.
Substituting it into equation (\ref{eq:R_d}), we get $R_d/r_{*}\propto t_\mathrm{inj}^{-4}\upsilon_\mathrm{fs}^{2}$, which is sensitive to $t_{\rm inj}$. For the same parameter set $M_*=15M_\odot,\,f_\mathrm{inj}=0.8,\,E_\mathrm{ej}=10^{51}\,{\rm erg}$,
%Using $R_d/r_*(t_\mathrm{inj}=10\,{\rm yrs})=0.21$, 
we find that $\tilde t_{\rm rise}$ reaches a maximum of $\sim 30$ days at $t_\mathrm{inj}\sim 7\,{\rm years}$.

From Figure \ref{fig:15Msun_lc_51erg}, the LC is found to decline more slowly for larger $t_{\rm inj}$. This is simply because the dense part of the CSM can be more extended for larger $t_{\rm inj}$, and the SN ejecta takes longer time to sweep up the entire CSM. The duration of this phase is roughly
\begin{eqnarray}
&&t_{\rm inj}\frac{\upsilon_w}{\upsilon_{\rm fs}} \nonumber \\
&\sim& 70\ {\rm days}\left(\frac{t_{\rm inj}}{10\ {\rm yrs}}\right)\left(\frac{\upsilon_w}{100\ {\rm km\ s^{-1}}}\right)\left(\frac{\upsilon_{\rm fs}}{5000\ {\rm km\ s^{-1}}}\right)^{-1},
\end{eqnarray}
which is clearly seen as a plateau with a cutoff in the case where $r_*/R_d\gg 1$ is achieved, e.g. for the $t_{\rm inj}=10$ yr and $30$ yr models. Otherwise the LC instead shows a monotonic decrease because the shock breakout component, with time dependence in equation (\ref{eq:breakout_flux}), would be dominant throughout the time the forward shock runs through the dense part of the CSM.

\subsubsection{Dependence on the Injected Energy $f_{\rm inj}$}
\begin{figure*}[t]
\centering
\includegraphics[width=\linewidth]{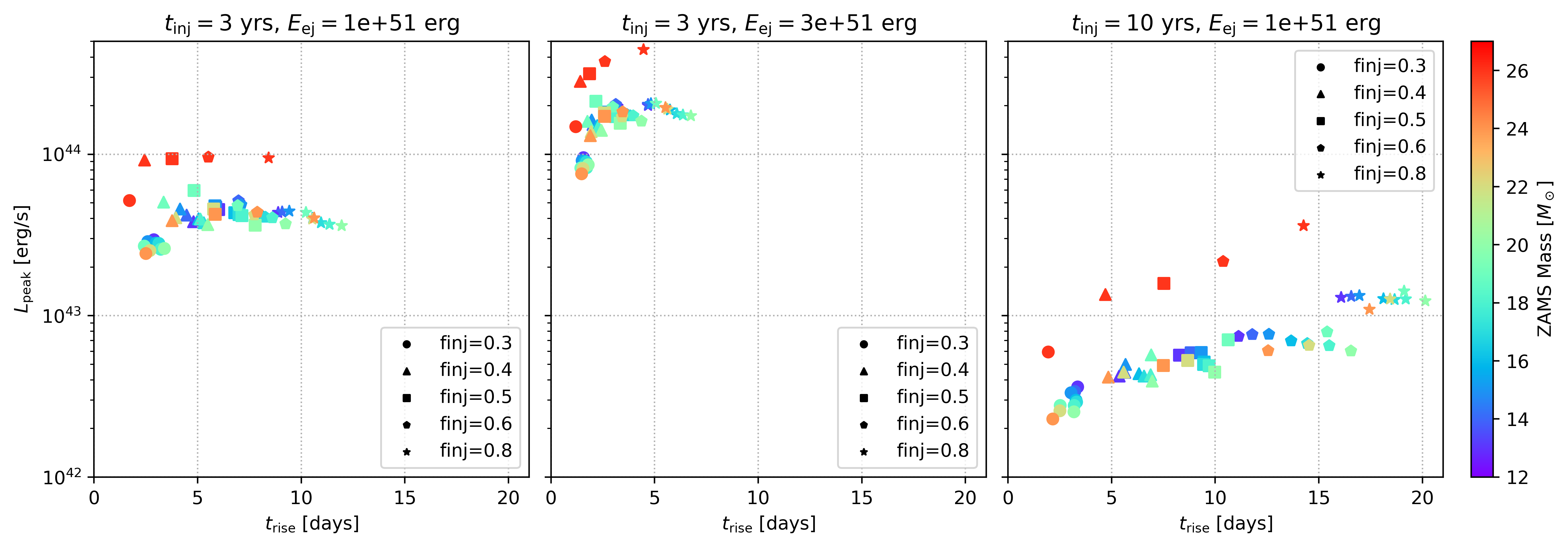}
\caption{The same as Figure \ref{fig:tinj_cmp}, but the dependence on $f_\mathrm{inj}$.}
\label{fig:finj_cmp}
\end{figure*}
In Figure \ref{fig:finj_cmp}, we plot the $t_\mathrm{rise}$--$L_\mathrm{peak}$ relation for $f_\mathrm{inj}$.
The interpretation is rather simple compared to section \ref{sec:tinj}.
More mass is erupted from larger injected energy $f_\mathrm{inj}$.
Except for the case of $M=19M_\odot,\,t_\mathrm{inj}=3\,{\rm yrs},\,E_\mathrm{ej}=10^{51}\,{\rm erg}$, $L_\mathrm{peak}$ for all models plotted in this figure monotonically increase with $f_\mathrm{inj}$.
This is roughly consistent with the previous result that the ratio of the radiated energy to the explosion energy is $\sim M_\mathrm{CSM}/(M_\mathrm{ej}+M_\mathrm{CSM})$ \citep{van_Marle_et_al_2010}.

The rise time $t_\mathrm{rise}$  also  monotonically increases with increasing $f_\mathrm{inj}$ because the erupted mass increases with $f_\mathrm{inj}$ and forms a CSM with a large optical depth.

\subsubsection{Dependence on the Explosion Energy $E_{\rm ej}$}
\begin{figure*}[t]
\centering
\includegraphics[width=\linewidth]{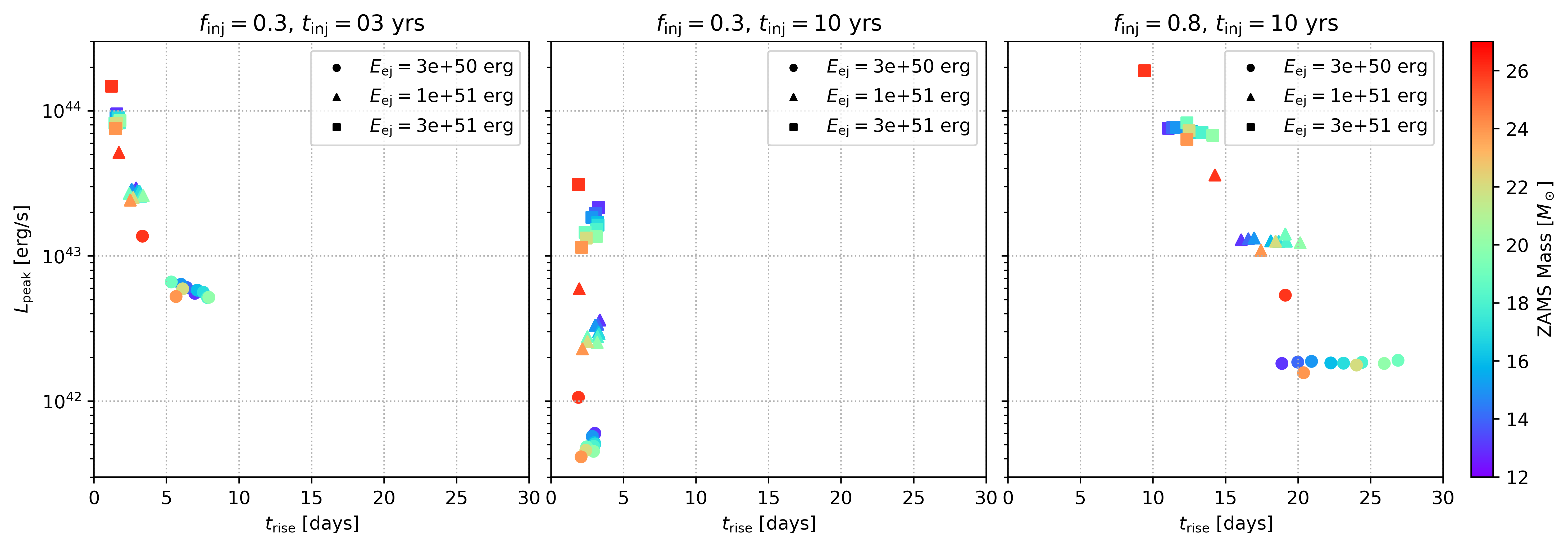}
\caption{The same as Figure \ref{fig:tinj_cmp}, but the dependence on $E_\mathrm{ej}$.}
\label{fig:Eexp_cmp}
\end{figure*}
The $t_\mathrm{rise}$--$L_\mathrm{peak}$ relations as a function of $E_\mathrm{ej}$ are plotted in Figure \ref{fig:Eexp_cmp}.
Since $\upsilon_\mathrm{fs}\propto E_\mathrm{ej}^{(n-3)/2(n-1.5)}$ for given $f_\mathrm{inj},\,t_\mathrm{inj},\,M_{*}$ \citep{Moriya_et_al_13}, it takes a shorter time to reach the peak luminosity for a larger explosion energy.
The larger $E_\mathrm{ej}$ gives higher peak to the LC.
It is expected from these results that the lower explosion energy gives lower peak and longer rise time.
%Since it is indicated from observations that the lower amount of $^{56}\mathrm{Ni}$ is created by the explosion compared to that of typical SNe \citep[e.g.,][]{Smith14b, Elias-Rosa18}, $E_\mathrm{ej}$ can be smaller than the typical explosion energy of $10^{51}\,{\rm erg}$. This low energy is consistent with the observed long rise time of some SNe IIn.

\subsubsection{Dependence on the ZAMS mass $M_*$}
Since the $t_\mathrm{rise}$--$L_\mathrm{peak}$ relation does not largely depend on $M_*$ except for $M_*=26M_\odot$ as seen in Figures \ref{fig:tinj_cmp}, \ref{fig:finj_cmp}, \ref{fig:Eexp_cmp}, we cannot solve the degeneracy only by the relation.
The difference of the $M_*=26M_\odot$ model is caused by the difference of the ejecta mass.
The calculated $M_\mathrm{ej}$ for the $M_*=26M_\odot$ model is $\approx5.4M_\odot$, which is much smaller than  $M_\mathrm{ej}\approx14M_\odot$ for the $M_*=17M_\odot$ model.
This difference arises from $M_{\rm rem}$, given in equation (\ref{eq:remnant_mass}) and plotted in Figure \ref{fig:progenitors}. The carbon-oxygen core mass of the $26M_\odot$ model exceeds the threshold of black hole formation, corresponding to the second case of equation (\ref{eq:remnant_mass}) where the value of $M_{\rm rem}$ jumps.
Since the ejecta with smaller mass moves fast for fixed $E_\mathrm{ej}$ ($\upsilon_\mathrm{ej}\propto M_\mathrm{ej}^{-1/2}$), the LC evolves rapidly and the peak luminosity becomes higher.

\subsubsection{Multi-band Light Curves}
\label{sec:multiband}
\begin{figure*}
\centering
\includegraphics[width=\linewidth]{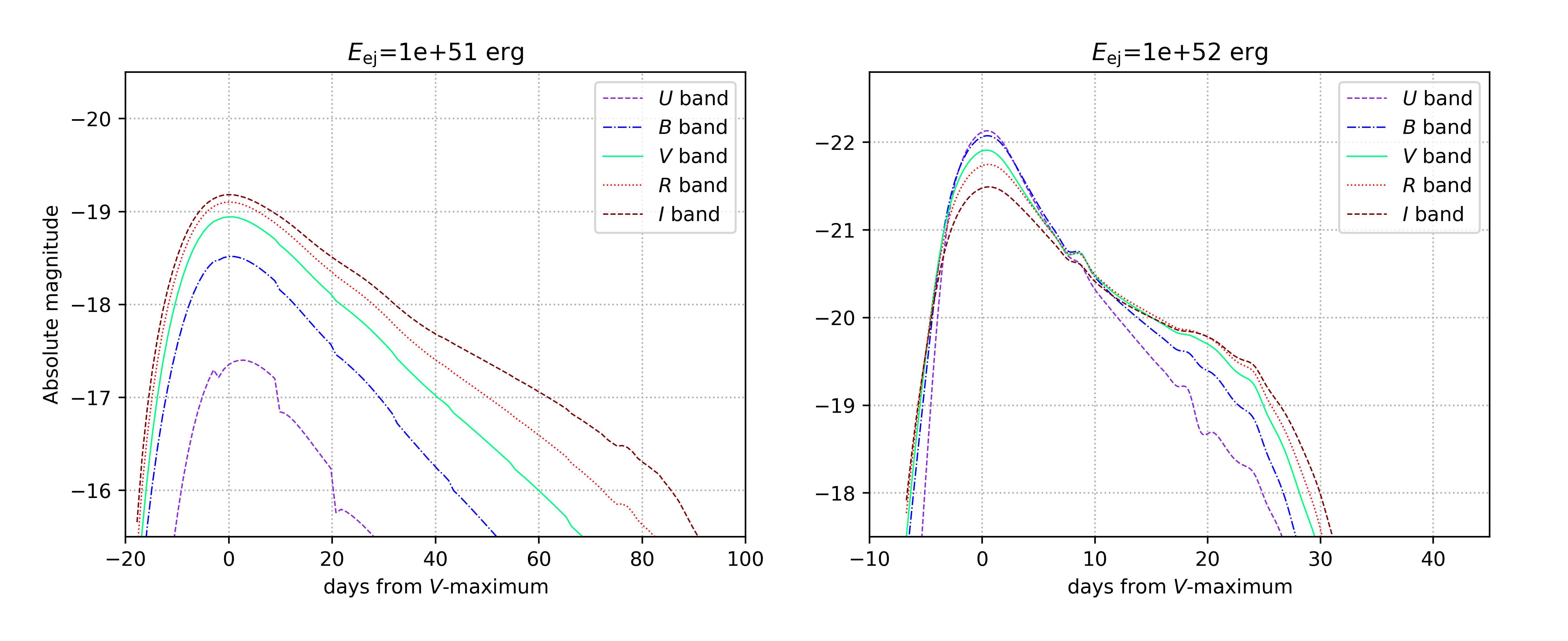}
\caption{$U,\,B,\,V,\,R,\,I$ magnitudes as functions of time. LCs with two different explosion energies are shown in different panels (see legends in each panel). Models with the following parameters are shown: $f_\mathrm{inj}=0.8,\,t_\mathrm{inj}=10\,{\rm yrs},\,M_{*}=15M_\odot$.}
\label{fig:multi_band_lc}
\end{figure*}
We plot in Figure \ref{fig:multi_band_lc} absolute magnitudes in $U,\,B,\,V,\,R,\,I$ bands.
As seen from absolute magnitudes of the model shown in the left panel, $U>B>V\gtrsim R\gtrsim I$ at around their peaks.
Since the color $V-R\sim0.1$ around the peak, the temperature of the model is estimated to be $\sim4,500\,{\rm K}$.
Changing $E_\mathrm{ej}$ from $10^{51}\,{\rm erg}$ to $10^{52}\,{\rm erg}$ with fixed other parameters, multi-band LCs are plotted in the right panel.
Compared to the model with lower explosion energy, the color becomes bluer (corresponding to the color temperature of $\sim14,000\ {\rm K}$.).
This is caused by the higher radiative flux than that of the lower explosion energy model.

Multi-band LC modelling can be a tool to additionally constraint the parameters jointly with the $t_\mathrm{rise}$--$L_\mathrm{peak}$ relation.

\subsubsection{Comparing with Observations}
\begin{figure*}
\centering
\includegraphics[width=\linewidth]{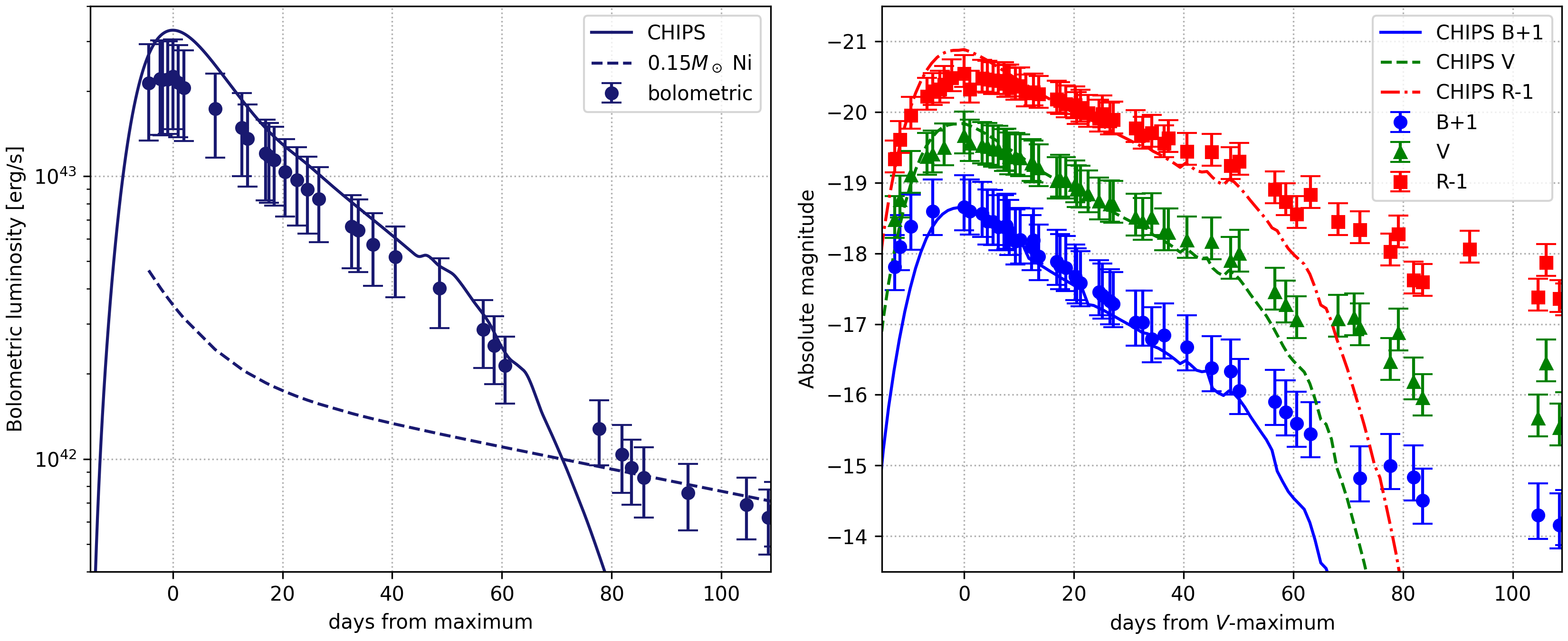}
\caption{Left panel: Comparison of the bolometric light curve of SN 1998S with that of a CHIPS model. The dashed line shows the luminosity of $^{56}$Ni and $^{56}$Co. Right panel: Comparison of the multi-band ($B,\,V,\,R$) light curves with those of a CHIPS model. The data are corrected for extinction using the color excess $E(B-V)=0.22^{+0.11}_{-0.08}$ \citep{Fassia2000MNRAS.318.1093F} and $A_B=4.1E(B-V),\,A_V=3.1E(B-V),\,A_R=2.32E(B-V)$ \citep{Savage_Mathis1979ARA&A..17...73S}.
A model with the following parameters is shown: $E_\mathrm{ej}=2.5\times10^{51}\,{\rm erg},\,f_\mathrm{inj}=0.7,\,t_\mathrm{inj}=11\,{\rm yrs},\,M_*=20M_\odot$.}
\label{fig:1998S_cmp_chips}
\end{figure*}
We compare CHIPS results with photometric data of SN IIn 1998S.
This SN was discovered on 1998 March 2.68 UT in NGC 3877 by Z. Wan \citep{1998IAUC.6829....1L}.
The nickel mass is estimated to be $0.15\pm0.05M_\odot$ from the tail of the bolometric light curve \citep{Fassia2000MNRAS.318.1093F}.
The left panel of Figure \ref{fig:1998S_cmp_chips} shows the bolometric LC of SN 1998S constructed by \citet{Fassia2000MNRAS.318.1093F} and the CHIPS result for parameters $E_\mathrm{ej}=2.5\times10^{51}\,{\rm erg},\,f_\mathrm{inj}=0.7,\,t_\mathrm{inj}=11\,{\rm yrs},\,M_*=20M_\odot$.
While we can successfully reproduce the bolometric LC for the first $\sim60$ days, the CHIPS model LC is fainter than the observed LC in the later epochs. This is because the energy input due to the radioactive decays of $^{56}$Ni and $^{56}$Co is not implemented in the CHIPS code (see also section \ref{sec:emission_ejecta}). The large explosion energy of $E_\mathrm{ej}=2.5\times10^{51}\,{\rm erg}$, which is necessary to reproduce the observed LC, is consistent with the production of the large nickel mass of $0.15M_\odot$ \citep[e.g.,][]{Hamuy2003ApJ...582..905H,Muller17}.
The multi-band LCs obtained by \citet{Fassia2000MNRAS.318.1093F,Liu2000A&AS..144..219L} are plotted in the right panel, together with those of the same CHIPS model.
As can be seen from the panel, each light curve in the early-middle epochs can be reproduced by CHIPS code as well.

In the following, we discuss the dependence of the characteristic observable quantities $L_\mathrm{peak},\,t_\mathrm{rise}$ on our results.
Thanks to the enormous observations of interaction-powered transients, we know the following features:
\begin{itemize}
    \item SNe IIn: $L_\mathrm{peak}\sim (\mathrm{a\, few}-10)\times 10^{42}\, {\rm erg/s}$, $t_\mathrm{rise}\sim (1-\mathrm{a\, few})\times10\, \mathrm{days}$ \citep{Ofek_et_al_2014}
    \item RETs: $L_\mathrm{peak}\sim (1-\mathrm{a\, few})\times 10^{43}\, {\rm erg/s}$, $t_\mathrm{rise}\sim (\mathrm{a\,  few}-10)\, \mathrm{days}$ \citep{Pursiainen_et_al_2018}
    \item FBOTs: $L_\mathrm{peak}\sim (1-\mathrm{a\, few})\times 10^{44}\, {\rm erg/s}$, $t_\mathrm{rise}\sim (1-\mathrm{a\, few})\, \mathrm{days}$ \citep{Arcavi_et_al_2016}
\end{itemize}
We can successfully obtain parameter spaces that are consistent the above observed regions in Figures \ref{fig:tinj_cmp}, \ref{fig:finj_cmp}, and \ref{fig:Eexp_cmp}.
The parameter spaces that give $t_\mathrm{inj},\,L_\mathrm{peak}$ of SNe IIn can be inferred as $f_\mathrm{inj}>0.8,\,E_\mathrm{ej}<10^{51}\,{\rm erg},\,t_\mathrm{inj}\sim 10\,{\rm yrs}$, for example. The indicated lower explosion energies are consistent with the small amount of $^{56}\mathrm{Ni}$ compared to that of typical SNe \citep[e.g.,][]{Smith14b, Elias-Rosa18}.
%To explain the characteristic observable of typical SNe IIn, the lower explosion energy and higher $f_\mathrm{inj}$ are required.
$t_\mathrm{inj}$ has to be selected so that $t_\mathrm{rise}$ takes a maximum for given other parameters.
Meanwhile, there is a degeneracy in parameters which give a shorter rise time and higher peak luminosity, typical observables of RETs and FBOTs.
Fitting the multi-color LCs to these observed quantities may break this degeneracy, which we will explore in a forthcoming study.

\section{Future Improvements}
\label{sec:future_work}
While we have shown that the current version of the CHIPS code successfully reproduces the diverse parameter spaces observed in interaction-powered SNe, there remain various simplifications and limitations. In this section we list our planned future improvements to the CHIPS code. We plan to release a new version of CHIPS when any of these updates have been done.

\subsection{Relaxing the Steady-state Assumption in the Shocked Region}
\label{sec:future_work_steadystate}
We resolve the shocked region at the shock rest frame with the assumption that the region is in a steady state.
However, with this assumption we cannot take into account the effect of the diffusion of photons in the shocked region at early epoch, which may affect the shock structure itself.
Moreover, we cannot resolve the shocked region at early times for some models in which the initial radius $r_0$ becomes larger than $2R_{*}$.

We plan to update our code by following the temporal evolution of the shocked region with time-dependent radiative transfer equation from the very early times. Then we can relax the approximations made when estimating the luminosity around CSM breakout (see Section \ref{sec:light_curve}), and we can predict the rise time and the peak luminosity more precisely.

\subsection{Inclusion of Emission from the Ejecta}\label{sec:emission_ejecta}
In the current version of the code we have taken into account only the emission powered by circumstellar interaction. While the interaction component is dominant around the peak of the LC, energy sources from the SN ejecta may become important when the CSM density drops at the late phase. In the future we plan to incorporate two robust energy sources into our LC model: internal energy deposited by the blast wave in the ejecta and radioactive decays of $^{56}$Ni and $^{56}$Co.

Though most of the internal energy in the ejecta is converted into kinetic energy as a star expands after the blast wave passes the stellar surface, a part of the internal energy is emitted from the expanding ejecta. The duration of this emission is determined by the expansion timescale, thus depends on the stellar radius. SNe originating from red supergiants shine for a few months, while those from Wolf-Rayet stars shine only for a few minutes by this energy source. If these SNe originate from red supergiants, then this source may contribute to the luminosity as much as another heating source, i.e., the radioactive decays of $^{56}$Ni and $^{56}$Co in the ejecta that emit energy with rates \citep{Arnett82}
\begin{eqnarray}
\epsilon_{\rm Ni} &\approx& 4.8\times 10^{10}\ {\rm erg\ g^{-1}\ s^{-1}}\exp[-t/8.8\,{\rm days}],\\
\epsilon_{\rm Co} &\approx& 2.6\times 10^8 \ {\rm erg\ g^{-1}\ s^{-1}}\exp[-t/110\,{\rm days}],
\end{eqnarray}
mainly in the form of $\gamma$-ray.
The energy from these two sources, once stored in the optically thick ejecta, is gradually released into space as hydrogen ions recombine from the outside to the inside of the ejecta in Type II-P SNe. This displays a plateau in the LC with luminosity $L_p$ lasting for $t_p$, given by the following formulae \citep{Popov93}
\begin{eqnarray}
L_p &\approx& 2\times 10^{42}\ {\rm erg\ s^{-1}}\left(\frac{\bar\kappa}{0.34\ {\rm cm^{2}g^{-1}}}\right)^{-1/3}\left(\frac{R_*}{500R_\odot}\right)^{2/3} \nonumber \\
&& \left(\frac{M_{\rm ej}}{10M_\odot}\right)^{-1/2}\left(\frac{E_{\rm ej}}{10^{51}{\rm erg}}\right)^{5/6} \left(\frac{T_\mathrm{ion}}{5054\ {\rm K}}\right)^{4/3}\\
t_p &\approx& 100\ {\rm days}\left(\frac{\bar\kappa}{0.34\ {\rm cm^{2}g^{-1}}}\right)^{1/6}\left(\frac{R_*}{500R_\odot}\right)^{1/6}, \nonumber \\
&& \left(\frac{M_{\rm ej}}{10M_\odot}\right)^{1/2}\left(\frac{E_{\rm ej}}{10^{51}{\rm erg}}\right)^{-1/6} \left(\frac{T_\mathrm{ion}}{5054\ {\rm K}}\right)^{-2/3},
\end{eqnarray}
respectively, where $T_\mathrm{ion}$ is the recombination temperature of hydrogen.
The ejecta are subject to engulfment and photoionization from the reverse shock, which can modify the resulting emission. Nonetheless this component more or less exists in SNe IIn as well, and a simple comparison implies it could be important for many of our LCs in the late phase. In fact SNe with co-existence of narrow hydrogen lines and plateau LC are observed \citep{Kankare12,Mauerhan13}, and sometimes classified as SN IIn-P \citep{Smith14b}. We note that the plateau would not affect our argument on the rise time and peak, since the former is generally much shorter than $t_p$.

%There are several upper limits on the synthesized nickel mass from observations of the late phase of SNe IIn, with values around $M_{\rm Ni}\lesssim 0.04M_\odot$ (e.g. \citealt{Smith14b,Elias-Rosa18}). Thus the input from radioactive decay is at most a few $\times 10^{48}\ {\rm erg}$. Unlike type I SNe, the contribution from radioactive decay is therefore not dominant.

There are a handful of inferences and upper limits on the synthesized nickel mass from late phase observations of SNe IIn. Most have values around $M_{\rm Ni}\lesssim 0.04M_\odot$ (e.g. \citealt{Smith14b,Elias-Rosa18}), while brighter SNe may have larger values (e.g. SN 1998S in the previous section having $M_{\rm Ni}\approx 0.15M_\odot$). The input from radioactive decay at the early phase is expected to be $10^{48}$--$10^{49}\ {\rm erg}$. Therefore unlike Type I SNe, the contribution from radioactive decay is not expected to be dominant in the early phase.

\subsection{Extension to Black Hole Formation}\label{sec:bh}
As mentioned in Section \ref{sec:technical_details}, not all the stars are expected to explode as canonical SNe with energy around $10^{51}$ erg. There are multiple tentative evidence from observations that a non-negligible fraction of massive stars die without being visible as canonical SNe and form BHs (e.g. \citealt{Smartt09,Horiuchi11,Kochanek14}).

However the mass ejection of these kinds of core-collapse is poorly known, and can have a diversity. A weak ``explosion" by reduction of gravity at the core due to neutrino emission \citep{Nadyozhin80} has been studied by recent simulations, which found explosions of energy $10^{46}$--$10^{48}$ ergs \citep{Lovegrove13,Fernandez18,Tsuna20,Ivanov21}. This can be significantly modified if an accretion disk can form around the nascent black hole, which are predicted to launch outflows and/or jets that can carry much more energy \citep{Woosley93,MacFadyen99,Kashiyama15,Quataert19,Tsuna21_at2018lqh}.

The mass eruption studied here can naturally occur for massive stars that form black holes as well. There is a recent claim that circumstellar interaction after black hole formation can explain some peculiar transients found in recent optical surveys \citep{Tsuna20}. Modelling these kinds of emission would be an important extension to our code.

\section{Conclusion}
\label{sec:conclusion}
We have developed the open source code CHIPS for modelling the LC of interaction-powered transients including the simulation of the mass eruption prior to a SN event.
CHIPS successfully calculates the LC for four key parameters, the ZAMS mass of the progenitor $M_*$, the injected energy normalized by the envelope's binding energy $f_\mathrm{inj}$, the time from energy injection to core-collapse $t_\mathrm{inj}$, and the explosion energy $E_\mathrm{ej}$.
After showing the methodologies of CHIPS, we explored the dependence of the resultant CSM and LCs on these parameters, $f_\mathrm{inj}=0.3,\,0.4,\,0.5,\,0.6,\,0.8$, $t_\mathrm{inj}=3,\,10,\,30\,{\rm yrs}$, $E_\mathrm{ej}=10^{51},\,3\times10^{51},\,10^{52}\,{\rm erg}$, $M_{*}=13,\,14,\,\cdots,\,26M_\odot$.
It is found that the peak luminosity $L_\mathrm{peak}$ becomes high for shorter $t_\mathrm{inj}$ and larger $E_\mathrm{ej}$.
Additionally, the rise time of a LC, $t_\mathrm{rise}$, is longer for larger $f_\mathrm{inj},\,E_\mathrm{ej}$.
In contrast to the simple relation of $L_\mathrm{peak}$ with $t_\mathrm{inj},\,E_\mathrm{ej}$, the dependence of $t_\mathrm{rise}$ on $t_\mathrm{inj}$ is complicated; a larger $t_{\rm inj}$ can make the extent of the CSM larger but also reduces the diffusion time (see Section \ref{sec:tinj} for details). For an optically thick CSM, there is a maximum value of $t_\mathrm{rise}$ when $t_{\rm inj}$ is varied, for given $f_\mathrm{inj},\,E_\mathrm{ej},\,M_{*}$.

We find that the parameter space we explore using CHIPS covers a wide range of $t_\mathrm{rise}$--$L_\mathrm{peak}$ space, including that of SNe IIn, RETs, and FBOTs.
We expect that CHIPS will serve as the tool for observers to utilize in obtaining physical quantities of interaction-powered transients.

CHIPS will be updated and released when we achieve any of the planned updates: i) to follow the temporal evolution of the shocked region between SN ejecta and CSM including radiation transport, ii) to solve radiative transfer equation in the unshocked ejecta in order to reproduce the LC at late phase, iii) to extend to the cases of black hole formation, or iv) any other improvements in the modelling that would be beneficial to the transient community.

\begin{acknowledgements}
We deeply thank the anonymous referee for helpful comments that greatly improved the manuscript, and Viktoriya Morozova for helpful guidance on using the SNEC code.
Y.T. is supported by the RIKEN Junior Research Associate Program. D.T. is supported by the Advanced Leading Graduate Course for Photon Science (ALPS) at the University of Tokyo, and by the JSPS Overseas Challenge Program for Young Researchers. This work is also supported by JSPS KAKENHI Grant Numbers 21J13957, JP19J21578, JP20H05639, MEXT, Japan.
\end{acknowledgements}

\software{MESA \citep[v12778;][]{Paxton11,Paxton13,Paxton15,Paxton18,Paxton19}, SNEC \citep[v1.01;][]{SNEC1}, Python libraries: Matplotlib \citep[v3.4.2;][]{Hunter:2007}, Numpy \citep[v1.17.4;][]{harris2020array}}

\newpage
\begin{appendix}
\section{Steps for Executing CHIPS}
\label{sec:CHIPS_execution}

\begin{figure*}
\centering
\includegraphics[width=\linewidth]{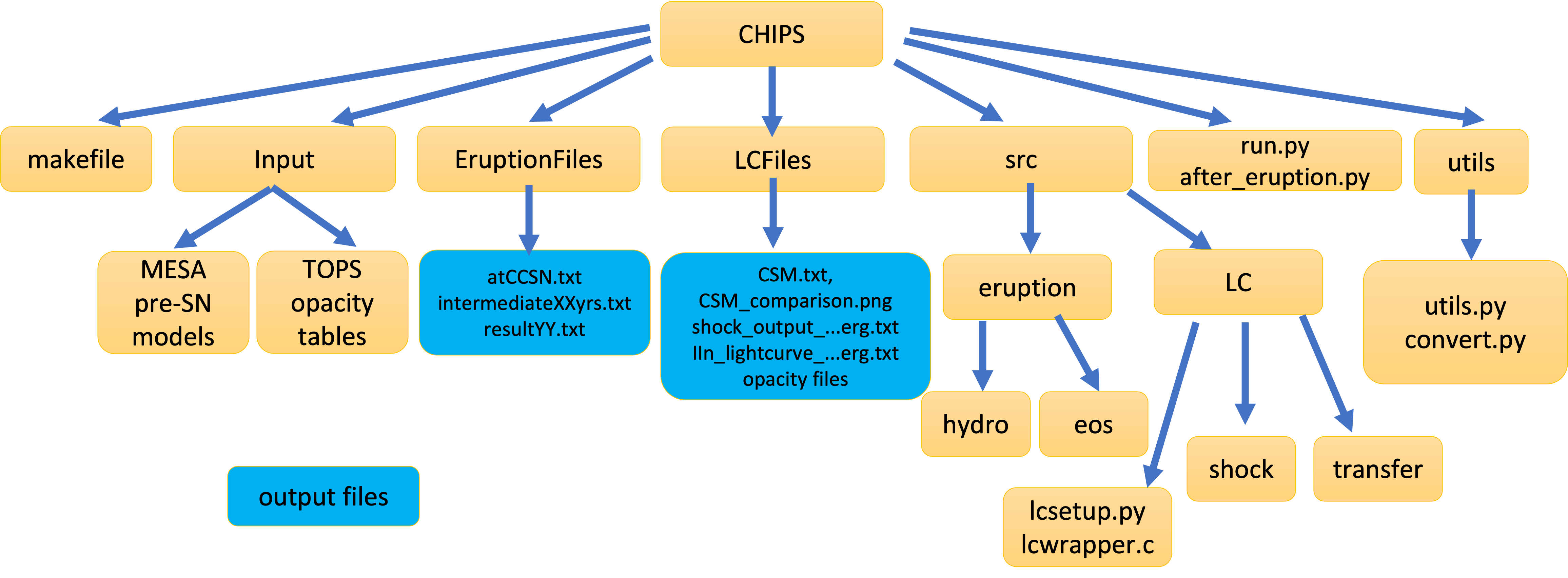}
\caption{Directory tree of CHIPS.}
\label{fig:chips_dir_fig}
\end{figure*}

\begin{table}[h]
 \caption{Description of output files. The units for dimensional quantities are all in CGS, except for $t$ in days and enclosed mass in $M_\odot$.}
 \label{table:file_description}
 \centering
  \begin{tabular}{c|cccccccc}
   \hline\hline
   File name & column 1 & 2 & 3 & 4 & 5 & 6 & 7 & 8 \\
   \hline
   intermediateXXyrs.txt & & enclosed mass & $r$ & $\upsilon$ & $\rho$ & $X$ & $Y$ & $P$\\
   shock\_output\_...erg.txt & $t$ & $u_\mathrm{rs}$ & $u_\mathrm{fs}$ & $r_\mathrm{rs}$ & $r_\mathrm{fs}$ & $F_\mathrm{fs}$ & $E_\mathrm{fs}$ & $\rho_\mathrm{fs}$\\
   IIn\_lightcurve\_...erg.txt & $t$ & $L(r=r_\mathrm{out})$ & -- & -- &--&--&--&--\\
   IIn\_lightcurve\_...erg\_mag.txt & $t$ & $U$ & $B$ & $V$ & $R$ & $I$ &--&--\\
   \hline
  \end{tabular}
\end{table}
Here we present the procedures to run the CHIPS code and describe the output files generated by the code. The name of output files and physical quantities listed in each file are shown in Table \ref{table:file_description}, and Figure \ref{fig:chips_dir_fig} explains the directory tree of CHIPS.
\begin{enumerate}
    \item First check whether gcc, gfortran, python, numpy and scipy can be called/imported on the command line. If any of these are not installed, install them.
    \item Then install mesa\_reader, a module that lets us easily extract data from the mesa output. This is used in some of our Python scripts. The detailed description of the module is in \href{http://mesa.sourceforge.net/output.html}{http://mesa.sourceforge.net/output.html}.
    \item Clone the CHIPS repository using the command  \begin{lstlisting}[basicstyle=\small\ttfamily, escapechar=\@, breaklines=true]
    git clone https://github.com/DTsuna/CHIPS.git 
    \end{lstlisting}
    \item The sample pre-SN models generated by the authors are compressed in a zip file in the directory \verb|input/mesa_models|. If one plans to use them, unzip this zip file. 
    \item At the top directory, compile the scripts for the mass eruption and light curve (LC) calculations using the makefile, with the command \verb|make|.
    \item Execute the script \verb|run.py| with the parameters in Table \ref{table:Parameters} as arguments. For example, to simulate a star with ZAMS mass 15 Msun and solar metallicity (assumed to be 0.014) which experiences mass eruption with parameters $f_{\rm inj}=0.3$ and $t_{\rm inj}=5$yr, and finally explodes with energy $E_{\rm ej}=10^{51}$ erg, run the following command
    \begin{lstlisting}[basicstyle=\small\ttfamily, escapechar=\@, breaklines=true]
    python run.py --tinj 5 --finj 0.3 --Eej 1e51 --analytical-CSM
        --stellar-model input/mesa_models/15Msun_Z0.014_preccsn.data 
    \end{lstlisting}
    The argument \verb|--stellar-model| specifies the MESA stellar model file to be used as input for the mass eruption calculation. The option \verb|--analytical-CSM| corrects the artificial shock compressions that arise from the mass eruption code before obtaining the LCs. For details of this procedure, see Section \ref{sec:mass_eruption} and Figure \ref{fig:csmremesh}.
\end{enumerate}

The mass eruption calculation reads in the MESA stellar model file given by the argument \verb|--stellar-model|, and outputs files with the CSM profile under the directory EruptionFiles/. The calculation also provides profiles at the intermediate stages, outputting one file per year with the name \verb|intermediateXXyr.txt|, where \verb|XX| are integer coded as two digit from 01 to $t_{\rm inj}$. Another set of files \verb|resultYY.txt| (YY being integers from 01 to 99), which finely resolves the period before and around the mass ejection event, is also created. In these files, the radius, velocity, density, hydrogen and helium fraction, and pressure of CSM XX years after the energy injection are listed as functions of the enclosed mass. The final CSM profile at $t_{\rm inj}$ is outputted with the name \verb|atCCSN.txt|, and is used to calculate the LCs.

The LCs of interaction-powered SNe are then simulated, with a set of explosion energies given by the argument \verb|--Eej|. This argument can be given multiple times to calculate LCs for multiple values of $E_{\rm ej}$, e.g., \verb|--Eej 1e51 --Eej 3e51|. When nothing is given, $E_\mathrm{ej}=\{1,\,3,\,10\}\times10^{51}\,{\rm erg}$ are used as default values. The LC calculation yields files named \verb|shock_output_...erg.txt| and \verb|IIn_lightcurve_...erg.txt| as outputs under the directory LCFiles/.
In \verb|LCFiles/shock_output_...erg.txt|, the radii of the reverse and forward shocks, the velocities of the both shocks, the radiative flux at the forward shock are listed as functions of time since explosion, while \verb|LCFiles/IIn_lightcurve_...erg.txt| list the luminosity at the edge of CSM $r=r_\mathrm{out}=3\times 10^{16}$cm, the photospheric radius, and the color temperature. Multi-band LCs (see examples in Figure \ref{fig:multi_band_lc}) can also be requested by the argument \verb|--calc-multiband|; in this case, another set of files called \verb|LCFiles/IIn_lightcurve_...erg_mag.txt| are created. These files contain LCs in U, B, V, R and I bands, with 1 day interval.

As done in the above example command, one can use the sample MESA models generated by the authors for the argument \verb|--stellar-model|.
We have 11 models for stars of a solar metallicity in the ZAMS mass range $13-26M_\odot$, with 8 models in $1M_\odot$ increments for masses up to $20M_\odot$ and 3 models in $2M_\odot$ increments for masses greater than $20\,M_\odot$. The pre-SN models are in a zip file in the directory input/mesa\_models/. After un-zipping this file, you will find MESA data files with the naming showing the mass and metallicity at ZAMS.

If a mass eruption calculation has already been done and the CSM profiles EruptionFiles/intermediateXXyrs.txt exist, one can use those files as CSM to obtain the LCs of interaction-powered SNe. This is done with the code \verb|after_eruption.py|, with a command like
\begin{lstlisting}[basicstyle=\small\ttfamily, escapechar=\@, breaklines=true]
    python after_eruption.py --Eej 1e51 --analytical-CSM 
        --stellar-model input/mesa_models/15Msun_Z0.014_preccsn.data
        --profile-at-cc EruptionFiles/intermediate05yr.txt 
\end{lstlisting}
By using this code one can avoid re-doing the mass eruption calculation, which is computationally costly compared to the LC calculation.

\section{Comparing with Other Works}
\label{sec:cmp_with_snec}
\begin{figure}
\centering
\includegraphics[width=\linewidth]{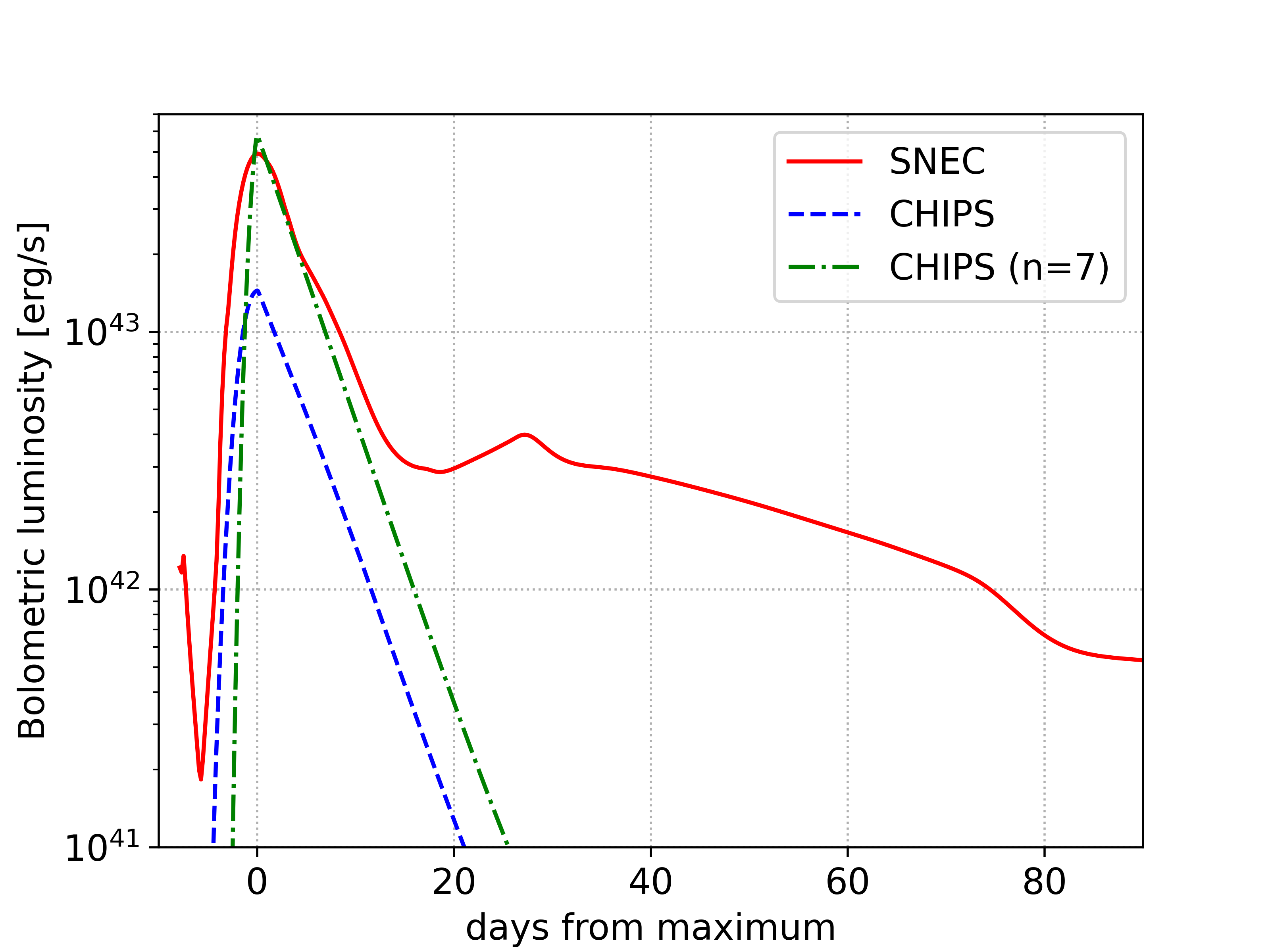}
\caption{Comparison of CHIPS results with SNEC code. Models with the following parameters $M_*=15M_{\odot},\,E_{\rm ej}=10^{51}\,{\rm erg},\,f_{\rm inj}=0.3,\,t_{\rm inj}=5\,{\rm yrs}$ are shown. The outer exponent of the ejecta $n=11$ is obtained by the equation (\ref{eq:npol_n}) (blue dashed line) while $n=7$ is given by hand (green dashdot line).}
\label{fig:snec_cmp}
\end{figure}
We compare our CHIPS code with past works on similar LC modelling, and adopt SNEC as an example as it is one of the most widely used open-source codes. SNEC\footnote{\href{https://stellarcollapse.org/index.php/SNEC.html}{https://stellarcollapse.org/index.php/SNEC.html}} is an open-source 1-D Lagrangian code for the hydrodynamics and equilibrium-diffusion radiation transport, aiming at modeling LCs of core-collapse SNe \citep{SNEC1}. More recently SNEC has also been used to simulate LCs of Type II-P/L SNe interacting with dense CSM \citep[e.g.,][]{SNEC2,Das_Ray_2017,Morozova20,Li21}.

We simulate an explosion and the resulting LC of our $15\ M_\odot$ progenitor using SNEC, using the CSM corresponding to a mass eruption with parameters $f_\mathrm{inj}=0.3,\,t_\mathrm{inj}=5\,{\rm yrs}$. We first stitch the profile of the inner helium core (obtained by MESA) to that of the outer hydrogen envelope obtained from the hydrodynamical simulation in the CHIPS code. Then we remove the innermost $\approx 1.4\ M_\odot$ from the computational region, corresponding to the value of $M_{\rm rem}$ in equation (\ref{eq:remnant_mass}) for our $15\ M_\odot$ progenitor. To simulate the explosion, SNEC injects energy by hand at the innermost computational region. Using the thermal bomb formalism, we set the value of the injected energy so that the final explosion energy becomes $E_\mathrm{ej}=10^{51}\,{\rm erg}$.

The red solid line in Figure \ref{fig:snec_cmp} shows the resulting LC obtained by extracting the luminosity at the outermost cell, which is always optically thin and travels much slower than the speed of light. The LC consists of two main phases: CSM breakout and subsequent CSM interaction ($t\lesssim10\,{\rm days}$), and the plateau phase powered by cooling and radioactive decay of $^{56}{\rm Ni}$ in the ejecta. In the same figure we also show in blue dashed lines the LC obtained by CHIPS code, which at present only simulates the contribution from the former CSM interaction. While the timescale of the emission is roughly in agreement, the peak luminosity in our calculation is dimmer by a factor of few. We presume this to be due to our simplified assumption of the ejecta density profile as a double power-law (equation \ref{eq:rho_ej}), where the outer ejecta follow $n\approx 11$ for this parameter set (equation \ref{eq:npol_n}). This simplified profile is often used in the literature, and in our case is required for our semi-analytical evaluation of the shock breakout emission in and after equation (\ref{eq:r_sh_diff}).

The actual ejecta density profile is predicted to have a smooth transition from the inner and outer asymptotic power-law indices \citep[][Figure 7]{Matzner99}. For interaction-powered SNe with massive CSM, the outer unshocked ejecta at CSM breakout will have a flatter profile than the asymptotic profile at the highest velocity. Thus a flatter density profile of $n=7$ for the outer ejecta was assumed in some of the previous works \citep{Chevalier_Irwin_2011,Chevalier12,Svirski12}. This is indeed seen from inspection of the density profile of the unshocked ejecta in the SNEC calculations. With this in mind, we additionally calculate another model where the exponent of the outer density profile is set by hand to be $n=7$, with all the other parameters fixed. We show this LC as green dash-dotted lines in Figure \ref{fig:snec_cmp}. We find that a model with $n=7$ agrees well with the LC from SNEC.

mIn essence, assumptions in the density profile of the outer ejecta can somewhat affect the LC, and the simple double power-law profile should eventually be updated to a more realistic one including curvature. Since the only limitation is from the semi-analytic framework of the breakout emission (which requires a power-law ejecta profile for self-similarity), this can be realized once the early-phase LC modelling is replaced to a more rigorous radiation transfer simulation, as outlined in Section \ref{sec:future_work_steadystate}.
%while the peak luminosity of a model with $n=11$ is smaller than the SNEC model. This is because the outer ejecta with a shallower density profile has larger kinetic energy.
%We check that the outer exponent of the ejecta set in the SNEC model is smaller than 11, which indicates that the difference between CHIPS and SNEC comes from the exponent.
% it is computationally expensive for longer $t_\mathrm{inj}$. Although CHIPS cannot simulate the LCs for $t_\mathrm{inj}$ shorter than $\sim3$ yrs due to a numerical instability caused by the high density of CSM, our preliminary tests find that CHIPS is computationally much cheaper than SNEC for wait time of years or longer. Therefore, SNEC and our code are complementary. We are searching for a way to suppress this instability.

We note that past works using SNEC have mainly focused on modelling multi-color/bolometric LCs of Type IIP/L SNe interacting with a relatively compact CSM, that formed by mass eruptions which occur within months to a few years before the terminal explosion.
Although this situation is similar to our model, at present it is difficult to conduct our calculations under their setting. This is because at such short timescales (i) the CSM extends only to several times the progenitor radius, so the ejecta is not homologous during the interaction phase, (ii) the LC code suffers numerical instabilities at high densities during the interaction, and possibly (iii) the CSM profile deviates from the analytical profile assumed throughout this work. These will be addressed in future updates of CHIPS to consistently model the early-phase LCs.

\end{appendix}

\bibliography{CSM}{}
\bibliographystyle{aasjournal}



\end{document}